\documentclass{aa}
\usepackage{graphicx,psfig}
\usepackage{natbib}
\usepackage{txfonts}
\usepackage{xcolor}
\def\be{\begin{equation}}
\def\ee{\end{equation}}
\begin{document}
\title{Grain size segregation in debris discs}

\author{P. Thebault\inst{1}, Q. Kral\inst{1}, J.-C..Augereau\inst{2}}
\institute{LESIA-Observatoire de Paris, CNRS, UPMC Univ. Paris 06, Univ. Paris-Diderot, France
\and
Universit\'e Joseph Fourier/CNRS, LAOG, UMR5571, Grenoble, France
}

\offprints{P. Thebault} \mail{philippe.thebault@obspm.fr}
\date{Received; accepted} \titlerunning{Size segregation in debris discs}
\authorrunning{Thebault et al.}

\abstract
%
{In most debris discs, dust grain dynamics is strongly affected by stellar radiation pressure. Because this mechanism is size-dependent, we expect dust grains to be spatially segregated according to their sizes. However, because of the complex interplay between radiation pressure, grain processing by collisions, and dynamical perturbations, this spatial segregation of the particle size distribution (PSD) has proven difficult to investigate and quantify with numerical models.
}
%
{We propose to thoroughly investigate this problem by using a new-generation code that can handle some of the complex coupling between dynamical and collisional effects. We intend to explore how PSDs behave in both unperturbed discs at rest and in discs pertubed by planetary objects. 
}
{We used the DyCoSS code developed by Thebault(2012) to investigate the coupled effect of collisions, radiation pressure, and dynamical perturbations in systems that have reached a steady-state. We considered two setups: a narrow ring perturbed by an exterior planet, and an extended disc into which a planet is embedded. For both setups we considered an additional unperturbed case without a planet. We also investigate the effect of possible spatial size segregation on disc images at different wavelengths.}
{We find that PSDs are always spatially segregated. The only case for which the PSD follows a standard $dn \propto s^{-3.5}ds$ law is for an unperturbed narrow ring, but only within the parent-body ring itself. For all other configurations, the size distributions can strongly depart from such power laws and have steep spatial gradients. As an example, the geometrical cross-section of the disc is very rarely dominated by the smallest grains on bound orbits, as it is expected to be in standard PSDs in $s^{q}$ with $q\leq-3$. Although the exact profiles and spatial variations of PSDs are a complex function of the set-up that is considered, we are still able to derive some reliable results that will be useful for image-or-SED-fitting models of observed discs.}
 {}

\maketitle

\section{Introduction} \label{intro}

Debris discs have been detected around a substancial fraction of solar-type main-sequence stars, $\sim 20$\% according to the latest Herschel surveys \citep{eiro13} \footnote{These values should be regarded as lower estimates, as they are limited by present-day detection limits. As an example, these detection limits would not allow the detection of the dust from the solar system's asteroid and Kuiper belts by an observer at 10-20 pc \citep{eiro13}}. These discs are observed through the emission of (or light scattered by) short-lived small grains, which are likely produced from destructive collisions amongst larger, undetectable objects \citep[e.g.][]{kriv10}, implying that impact velocities are high and that these systems are probably dynamically stirred \citep{theb09,kenn10}. These characteristics, together with the complex spatial structures observed for most resolved discs, suggests that debris discs are part of planetary systems. The existence of such  long-lived collisionally active (and thus dust-producing) belts made of planet-formation remnants (such as the asteroid and Kuiper belts) is also supported by planet formation scenarios, which suggest that they probably are a relatively common outcome \citep[e.g.,][]{liss06}.

In most cases, dusty debris discs are the most easily observable counterparts of such planetary systems. And it has been the main goal of numerical debris disc models to use the observed dust as a tool for reconstructing these system's global structure, be it the vast population of large collisional progenitors or the presence of hidden planetary objects. This objective represents a difficult challenge for disc modellers, because of the complex interplay between all the processes at play in debris discs: mutual collisions, dynamical perturbations, stellar radiation and winds, etc. A major difficulty comes from the fact that impacts in debris discs are highly erosive, each of them expected to produce clouds of small fragments that are impossible to track down and individually follow in deterministic numerical codes. An additional challenge comes from the scale of the population that has to be modelled, from the micron-to-millimetre-sized observed grains to their planetesimal-like parent bodies as well as possible planet-sized perturbers.

For all these reasons, most previous modelling efforts have focused on studying separately either the collisional evolution of discs, with statistical codes focusing on these systems' particle size distribution (PSD), total mass evolution, and dust-induced excess luminosities \citep[e.g.][]{keny04,kriv06,theb07,wyat07,lohn08,shan11,gasp12}, or their dynamics, with $N$-body codes investigating, for instance, how the presence of stellar or planetary perturbers might explain observed spatial structures \citep[e.g.][]{moro02,kuch03,dell05,rech08,lest11}.

These investigations have yielded important results, but excluding the dynamics in statistical particle-in-a-box codes or collisions in $N$-body models limits the problems that they can quantitatively explore \citep[for a  discussion on these limitations, see for example][]{theb12a}.
To overcome these limitations, recent years have seen the first attempts to combine the dynamical and collisional modelling of debris discs. This new generation of codes has been used to investigate the signatures of planets and companion stars in collisional discs \citep{star09,kuch10,theb12a,theb12b} as well as the evolution of transient massive collisional events \citep{grig07,kral13}.

Another important question that these new codes allow to investigate, but which has remained relatively unexplored to this date, is that of the spatial distributions of the PSDs in debris discs. We indeed expect the interplay between gravitational dynamics, radiation pressure, and collisions to induce size-dependent trends in the way dust grains are spatially distributed. The main reason for this is that stellar radiation pressure creates a size-dependence in the dynamical evolution, and thus the spatial distribution, of small grains, which in turn creates spatial variations of the collisional evolution because of local (and size-dependent) variations in terms of collision rates and impact velocities. As a matter of fact, it is already known from collisional statistical models that even in non-perturbed systems the size distribution of grains is not spatially homogeneous \citep{theb07}.

This question is of great importance, because different PSDs in different regions can have significant observational consequences, especially when imaging a disc at different wavelengths, since both scattered light and thermal emission strongly depend on the sizes of the emitting grains. Such information would be highly valuable for models desgined to fit images and SEDs (Spectral Energy Distributions), such as GRaTeR \citep{auge99} or SAnD \citep{erte12}, whose aim is to find a best match to observational constraints by testing large samples of synthetic discs obtained from a parametric exploration of grain size and spatial distributions. Given the sheer scale of the parameter space exploration inherent to such best-fit codes, a full exploration of all free parameters is indeed impossible, and some simplifying assumptions have to be made, in particular, that the PSD is the same everywhere in the system \citep[e.g.][]{lebr12,menn13}. This limitation could be relaxed if simple laws regarding the spatial distribution of PSDs could be implemented.

We propose here to numerically investigate the grain-size segregation in debris discs and its consequences on resolved images, using the DyCoSS code developed by \citet{theb12a}, which was designed to study systems at dynamical and collisional steady-state. We considered both unperturbed debris discs at rest and discs perturbed by one embedded or exterior planet.

\section{Numerical procedure}

\subsection{DyCoSS code}

We here only briefly recall the main characteristics of the algorithm, and refer to \citet{theb12a} and \citet{theb12b} for a detailed description.

DyCoSS ("Dynamics and Collisions at Steady-State") has an N-body structure to deterministically follow the spatial evolution of $N_{num}$ test particles, into which collision-related features are implemented.
We stress that DyCoSS, in contrast to the more sophisticated LIDT-DD code by \citet{kral13}, does not fully self-consistently couple collisions and dynamics. All test particles are assumed to be dusty collisional fragments produced, from parent bodies large enough to be unaffected by radiation pressure, following a size distribution in $dn/ds \propto s^{q}$. These particles are then assigned a size- and location-dependent collisional lifetime after which they are removed from the system. In this respect, it shares several characteristics with the CGA algorithm by \citet{star09}.

The main idea behind DyCoSS is that the dust present in the system at a given time $t_0$ consists of grains that have been collisionally produced at different epochs in the past. This means that, under the abovementioned assumption that particles have collisional removal times that can be estimated, the total dust population at a time $t_0$ can be reconstructed by running a series of runs, each for dust that has been produced at $t_0$, $t_0-dt$, $t_0-2dt$,...,$t_0-n.dt$, ... In the general case, this procedure would be impracticable because the number of runs that needs to be performed would be almost infinite. 
However, for the specific case of \emph{one} dynamical perturber and for a system at \emph{steady-state}, the system becomes periodic with respect to the perturber's orbital period $t_{orb}$, so that only a limited number of runs has to be performed, each for dust grains released at a different initial position of the perturber (here, the planet) on its orbit. In this case, the evolution of dust particles released at $t_0-\Delta t$, $t_0-\Delta t-t_{orb}$, $t_0-\Delta t-2t_{orb}$, ...,$t_0-\Delta t-n.t_{orb}$ can be retrieved from the same single run, started when the perturber was at the orbital position it had at $t_0-\Delta t$. If we divide the perturber's orbit into $N_{pos}$ segments, then, for each of the $N_{pos}$ runs, all particle positions have to be recorded at regularly spaced time intervals, corresponding to a fraction $t_{orb}/N_{pos}$ of the perturber's orbital period, until all particles have exceeded their collisional lifetimes. These collisional lifetimes are derived using surface density maps obtained from series of "parent-body" runs \citep[see ][, for further details]{theb12a}.

In practice, the procedure is divided in three steps: 1) For each of the $N_{pos}$ runs, a purely dynamical collisionless parent-body run is performed, following the evolution of seed particles only submitted to gravitational forces, until a dynamical steady-state is reached; 2) from each of these $N_{pos}$ steady-state parent-body discs, a population of dust grains is released following a $dn/ds \propto s^{q}$ size distribution, whose dynamical evolution is followed taking this time into account radiation pressure and collisional lifetimes. During this evolution, particles are progressively removed, either by dynamical ejection or collisions. The positions of all surviving grains are recorded every $t_{orb}/N_{pos}$ time intervals until no particle is left; 3) the stored particle positions for each of these collisional runs are then recombined to produce maps of the system's steady-state surface density at different positions of the perturber on its orbit.

We studied two main cases: unperturbed discs, and discs perturbed by one planet. As in \citet{theb12b}, we considered, for each case, two initial disc morphologies, that is, either an extended disc, or a narrow ring. For the cases with perturber, the planet is either embedded in the disc or exterior to it. The planet and disc are always assumed to be coplanar.

For the sake of clarity, and to facilitate comparing our results between cases with different stellar types, we parameterized particle sizes $s$ by $\beta$, the ratio of the stellar radiation pressure force to stellar gravity, under the assumption that $\beta=0.5\times s_{cut}/s$, where $s_{cut}$ is the size below which dust grains are blown out of the system (if produced from a parent body on a circular orbit). 
We considered grains of sizes between $\beta=0.5$ and $\beta=0.0125$ (that is from $s_{min}$=$s_{cut}$ to $s_{max}$=$40\times s_{cut}$). We assumed that these grains are collisionally produced from large parent bodies ($\beta=0$) following a size distribution\footnote{Note that this size distribution is not the one at collisional equilibrium but corresponds to the \emph{crushing law} for fragments produced after one fragmenting impact \citep[e.g.,][]{kuch10}}in $dn/ds \propto s^{-3.5}$. Given this steep size distribution, it is impossible to model the whole $0.0125\leq \beta\leq0.5$ size range with the typically $N\sim10^{5}$ number of particles of our $N$-body runs. We therefore divided each of the $N_{pos}$ runs into three subruns: a "small grains" run ($0.15\leq \beta\leq0.5$), an "intermediate grains" run ($0.05\leq \beta\leq0.15$), and a "large grains" run ($0.0125\leq \beta\leq0.05$). For the number $N_{pos}$ of different orbital positions of the perturbing planet that are considered, we took $N_{pos}=10$. This means that each planet-disc configuration simulation consists of $10\times3=30$ separate runs that are then recombined.

As explained in more detail in \citet{theb12a}, the absolute collisional lifetimes of grains are tuned in by one free parameter, that is, the average vertical optical depth $\tau_0$ in the disc \footnote{As mentioned earlier, the \emph{relative} collisional lifetimes between grains are then derived from their sizes and from surface density maps obtained from the parent-body runs}. We took as a reference value $\tau_0=2\times 10^{-3}$, typical for dense young debris discs such as $\beta$ Pictoris \citep{goli06}, but explored other values.

For the perturbing planet, we explored different values of its mass $m_p$ (as parameterized by its ratio $\mu$ to the stellar mass) and semi-major axis $a_p$. Since we are interested in identifying and deriving general trends for PSD segregation, we restricted our already large parameter exploration to the circular orbit case with $e_p=0$.

All main parameters for the simulations' setup are summarized in Table \ref{setup}.

\begin{table}
\begin{minipage}{\columnwidth}
\caption[]{Nominal set-up for the narrow-ring and extended-disc runs. Parameters denoted by an * are explored as free parameters.}
\renewcommand{\footnoterule}{}
\label{setup}
\begin{tabular*}{\columnwidth} {ll}
\hline
PARENT-BODY RUN &  \\
\,\,\,Number of test particles & $ N_{PB}=2 \times 10^{5}$\\
\,\,\,Initial radial extent & $30<r<130$\,AU (wide-disc)\\
\,\,\,& $45<r<53$\,AU (narrow-disc)\\
\,\,\,Initial eccentricity & $0\leq e \leq 0.01$\\
\,\,\,Initial surface density & $\Sigma \propto r^{-1}$\\
PERTURBING PLANET & \\
\,\,\,$m_p/m_*$ & $^{*}\mu=2\times 10^{-3}$\\
\,\,\,eccentricity& $e_p=0$\\
\,\,\,semi-major axis&$^{*} a_p=75\,$AU\\
COLLISIONAL RUNS & \\
\,\,\,Average optical depth & $^{*} \tau_0= 2\times10^{-3}\,$\\
\,\,\,Number of test particles & $ N_{num}=2 \times 10^{5}$\\
\,\,\,Size range\footnote{$s_{cut}$ is the radiation pressure blow-out size} & $s_{cut}\leq s \leq 40 s_{cut}$\\
\,\,\,Size distribution at $t=0$ \footnote{when released from the parent-body population} & $dn(s) \propto s^{-3.5}ds$\\
\hline
\end{tabular*}
\end{minipage}
\end{table}
\begin{table}
\begin{minipage}{\columnwidth}
\caption[]{List of all simulations performed for this study. The two nominal cases for the narrow ring and extended disc set-ups are marked by an *}
\label{runs}
\begin{tabular*}{\columnwidth} {c|c|c|c|c|c|}
\hline
RUN & $r_{min}$ (AU) & $r_{max}$ (AU) & $\mu$ & $a_p$ (AU)& $\tau_0$\\
\hline
A & 45 & 53 & $0$ & - & $2\times 10^{-3}$\\
B & 30 & 130 & $0$ & - & $2\times 10^{-3}$\\
C* & 45 & 53 & $2\times 10^{-3}$ & 75 & $2\times 10^{-3}$\\
D & 45 & 53 & $2\times 10^{-3}$ & 75 & $5\times 10^{-5}$\\
E & 45 & 53 & $2\times 10^{-3}$ & 75 & $2\times 10^{-2}$\\
F & 45 & 53 & $2\times 10^{-4}$ & 62 & $2\times 10^{-3}$\\
G* & 30 & 130 & $2\times 10^{-3}$ & 75 & $2\times 10^{-3}$\\
H & 30 & 130 & $2\times 10^{-3}$ & 75 & $5\times 10^{-5}$\\
I & 30 & 130 & $2\times 10^{-3}$ & 75 & $2\times 10^{-2}$\\
J & 30 & 130 & $2\times 10^{-4}$ & 75 & $2\times 10^{-3}$\\
\hline
\end{tabular*}
\end{minipage}
\end{table}

\section{Results}

\subsection{Unperturbed system}

It has been known for more than a decade that, even in an axisymmetric disc at rest, the spatial distribution of grains depends on their sizes. The simple parametric model of \citet{auge01} already pointed out that radial size segregation naturally occurs because of radiation pressure forces. For the specific case of an unperturbed narrow annulus, \citet{stru06} and \citet{theb08} have shown that the coupling of stellar radiation pressure and collisions in the ring results in a size segregation beyond the main ring, with particles of decreasing sizes at increasing distances to the ring. This creates a "natural" surface density profile in $r^{-1.5}$ beyond the ring after integrating over all particle sizes.

The situation for an unperturbed extended disc is more complex, but, in this case too, radial size segregation is a feature that can in principle be mapped with statistical collisional codes with a 1-D resolution, such as the ACE code developed in Jena \citep{kriv05,kriv06,mull10,vite10}. However, even if size segregation has indeed been obtained in several such studies \citep[e.g.,][]{theb07}, it was always as a side effect that was never thoroughly investigated in itself.   
We reinvestigate this issue here using the DyCoSS code instead of a statistical collisional code, because it will allow an easy comparison with the cases with a planet.

\subsubsection{Narrow ring} \label{narri}

\begin{figure*}
\makebox[\textwidth]{ 
\includegraphics[scale=0.5]{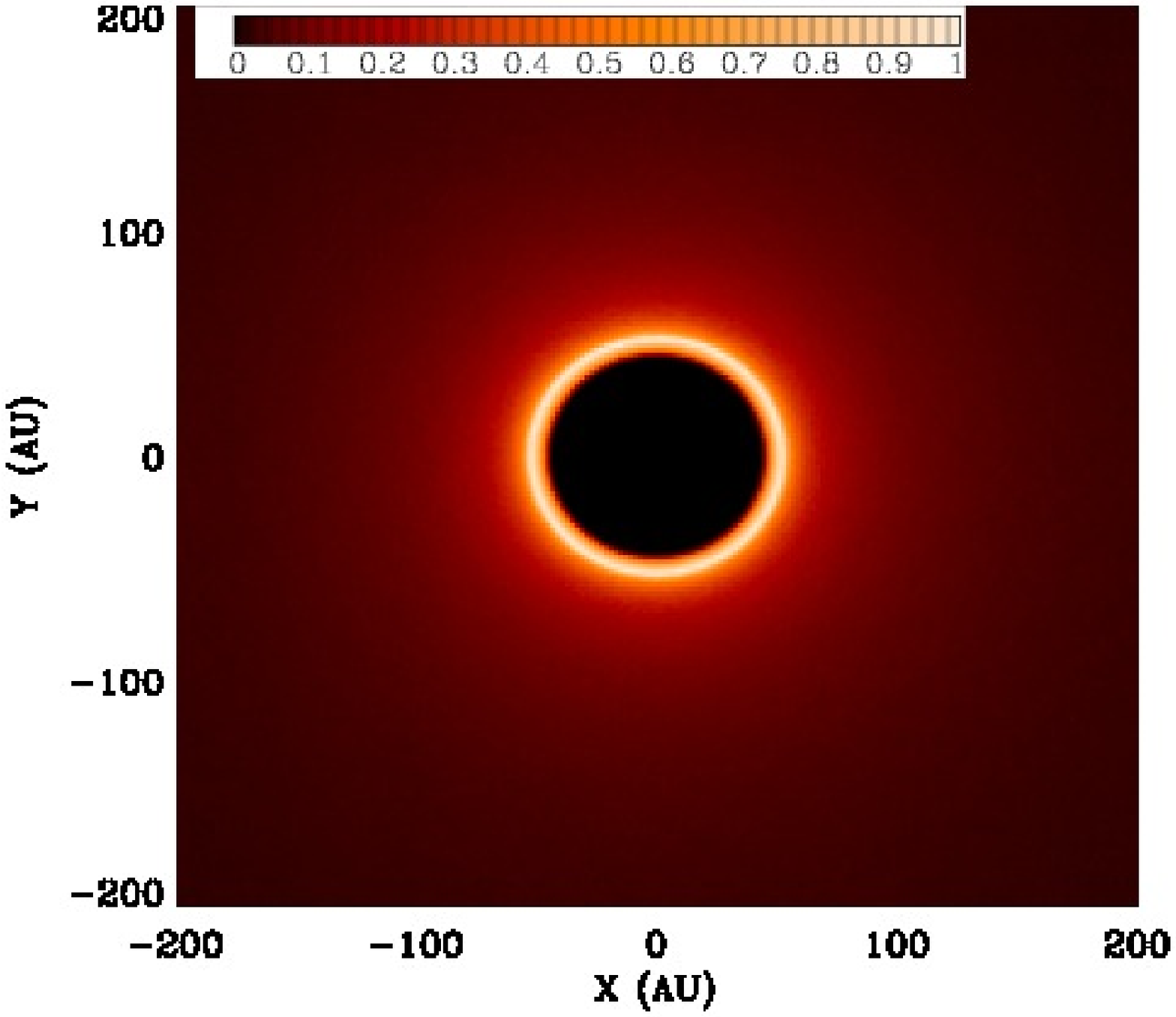}
\color{white} 
 \put(-200,40){a)} 
\color{black}
\includegraphics[scale=0.5]{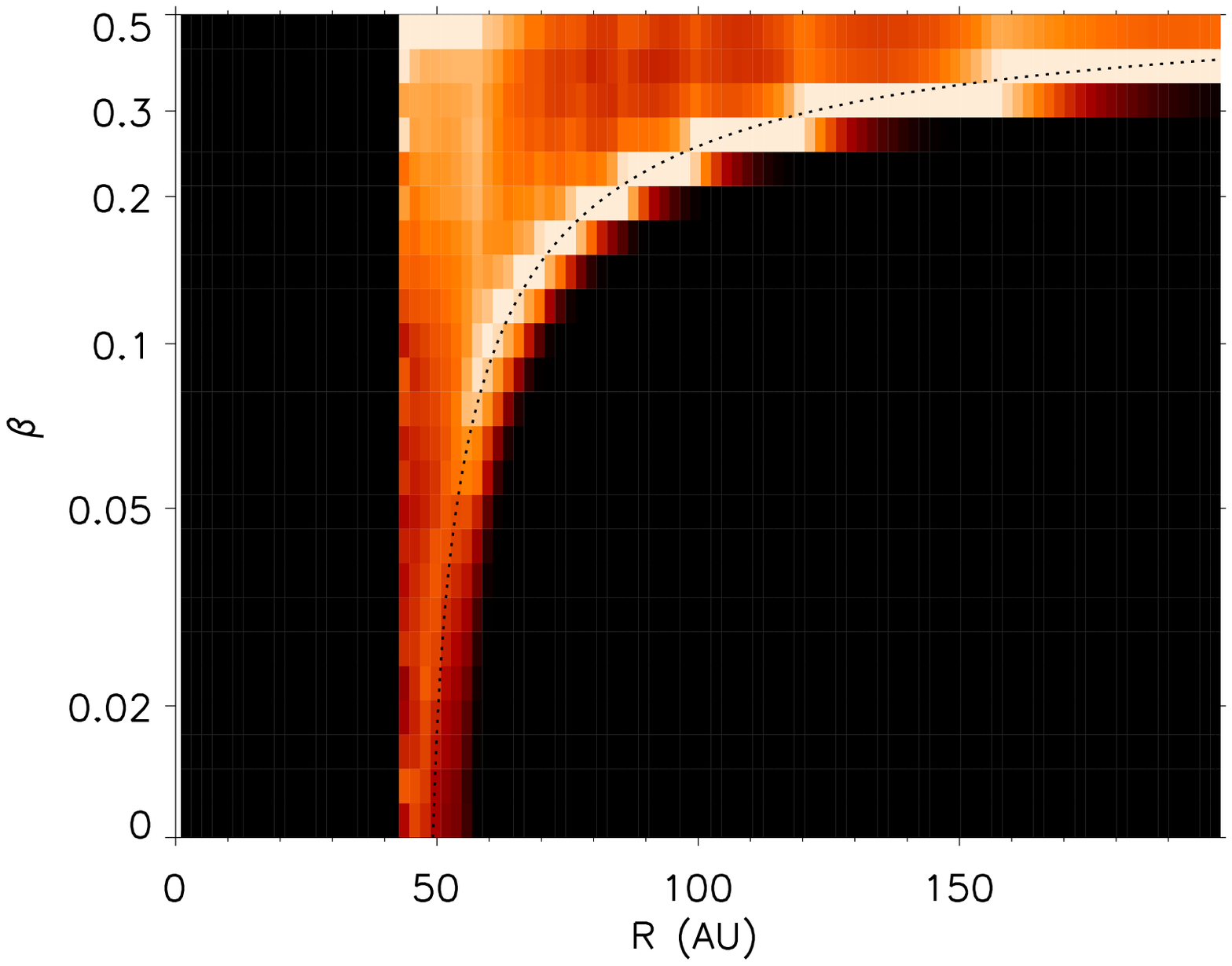}
\color{white} 
 \put(-50,40){b)} 
\color{black}
}
\caption[]{Steady-state for an unperturbed disc initially confined to a narrow birth ring between 45 and 53\,AU. \emph{a}): Normalized surface density profile viewed head-on. \emph{b}): Azimuthally averaged radial distribution of the geometrical cross-section $\sigma$ as a function of particle sizes (as parameterized by their $\beta$ value). At each radial distance, the $\sigma$ distribution has been normalized to 1. The dotted line shows, for a given radial distance $r$, the $\beta$ value of particles whose periastron is located in the middle of the ring and whose apastron is located at $r$.
}
\label{intno}
\end{figure*}

\begin{figure*}
\begin{minipage}{\columnwidth}
\centering
\includegraphics[width=\columnwidth]{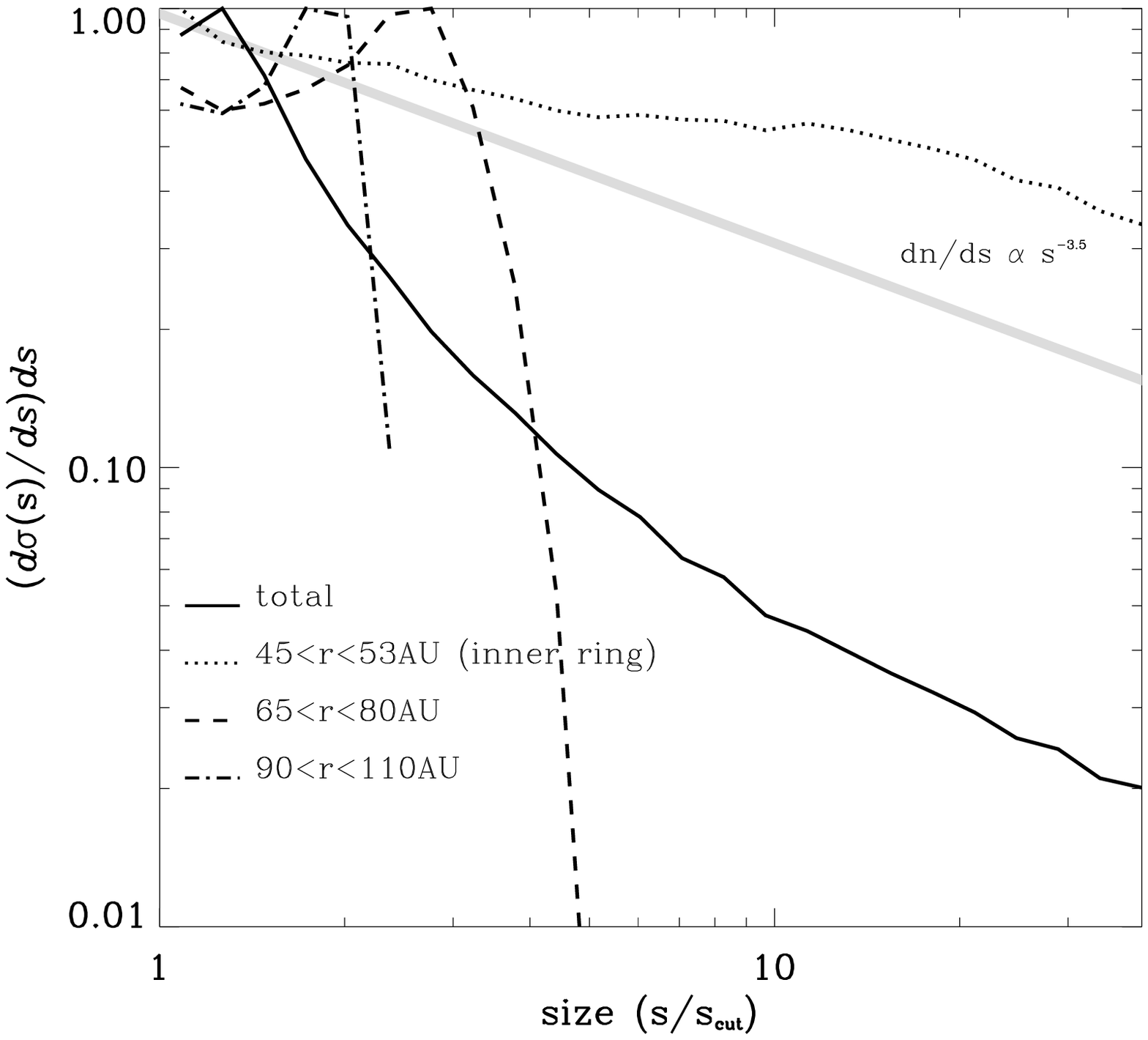}
\caption[]{Unperturbed narrow-ring case. Geometrical cross section distribution in each logarithmic size bin for several radial domains in the system. The distribution is normalized to 1 for each radial range.
}
\label{intnosz}
\end{minipage}%
\hfill
\begin{minipage}{\columnwidth}
\hfill
\includegraphics[width=\columnwidth]{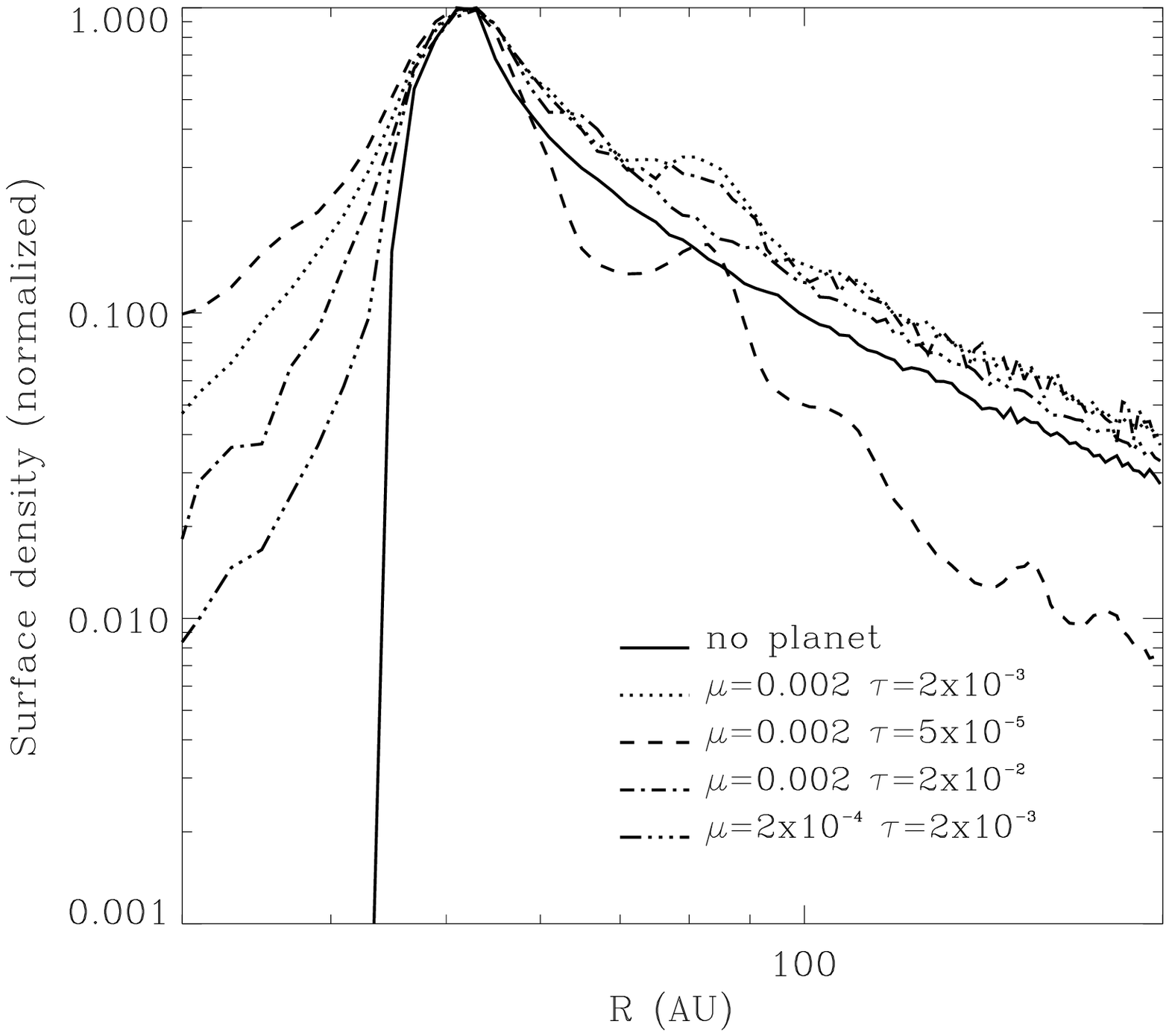}
\caption[]{Radial profile (normalized) of the vertical optical depth for different set-ups of the narrow-ring case. For the cases with planet, the profile is taken along a cut passing at 90 degrees from the planet's position. The planet is on a circular orbit and is located at 75AU for the nominal $\mu = 0.002$ cases, and at 62AU for the $\mu = 2\times10^{-4}$ one.
}
\label{intnocut}
\end{minipage}
\end{figure*}

We first present in Fig.\ref{intno} the results obtained for an unperturbed, initially confined narrow ring (run A, see Tab.\ref{runs}). The surface density map shows the well-known result that, at steady-state, there is dust well beyond the outer edge of the initial ring. This dust consists of small grains, collisionally produced in the ring and placed on high-$e$ orbits by radiation pressure. 

Figs.\ref{intno}b and \ref{intnosz} present in more details the shape of the size distribution (as represented by the geometrical cross section $\sigma$ per logarithmic size interval) within and beyond the ring. 
\emph{Within} the initial "birth" ring we retrieve a somehow counter-intuitive feature, identified by \citet{stru06} and \citet{theb08}: the size distribution settles into a profile relatively close to $dn/ds \propto s^{-3.5}$ even though small grains produced in the ring spend most of their time on high-$e$ orbits outside of it and would naively be expected to be underabundant in the birth ring. The -3.5 equilibrium law holds regardless, because these high-$e$ orbits pass through almost empty regions and small grains can only be destroyed when they re-enter the birth ring, so that they continue to accumulate until a steady-state is reached when their collisional production and destruction rates balance each other \emph{within} the birth ring \citep{theb08}. The consequence of this is a global overdensity of small grains when integrated over the whole system (full line curve in Fig.\ref{intnosz}), and an average surface density having a radial profile close to $r^{-1.5}$ beyond the ring (Fig.\ref{intnocut}).
Note, however, that since the regions beyond the ring are not totally empty (see below), some collisional destruction occurs there. This is the reason for the slight flattening of the ring-PSD in the $\lesssim 5s_{cut}$ domain, as well as the somehow steeper than $-1.5$ slope of the radial density profile in the outer regions.

Beyond the ring, the size distribution takes on a very different shape and becomes strongly peaked, around a value $s_{peak}(r)$ that decreases with increasing distance to the ring (Fig.\ref{intnosz}). Fig.\ref{intno}b shows that, at any radial location $r$ beyond the main ring, the geometrical cross section is dominated by the largest grains that can reach $r$ from a production location in the main ring,that is, grains whose periastron is located in the ring and whose apoastron is lcoated at $r$. This behaviour had been anticipated (without being demonstrated) by \citet{stru06} and arises because these grains are the ones that spend the longest time at the distance $r$.
In terms of the particle size that dominates the geometrical cross section, there is thus a sharp transition at the outer edge of the ring, where it jumps from $\sim s_{cut}$ to $\sim 5s_{cut}$ (Fig.\ref{intno}b).

\subsubsection{Extended disc}

\begin{figure*}
\makebox[\textwidth]{
\includegraphics[scale=0.5]{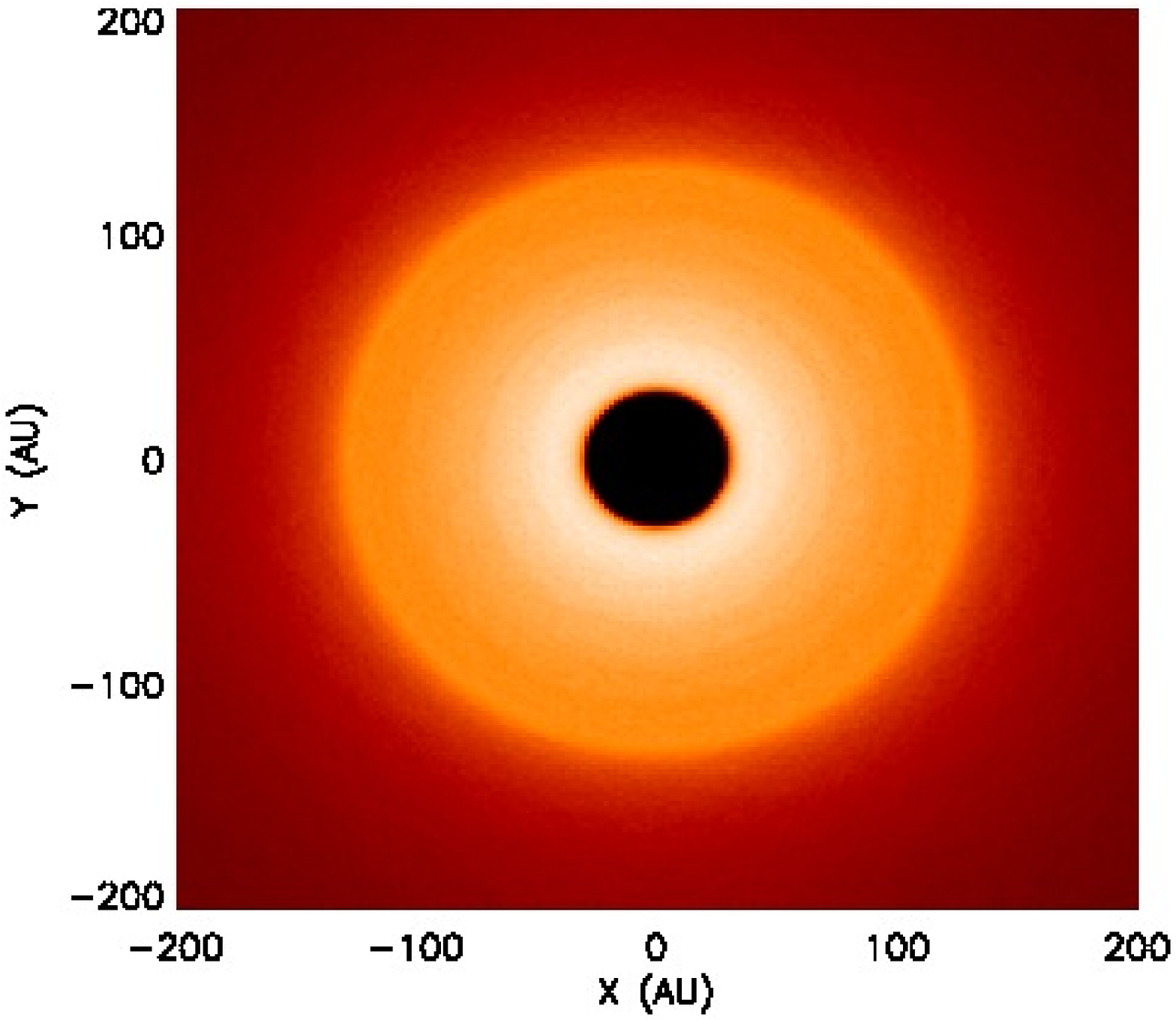}
\color{white} 
 \put(-200,40){a)} 
\color{black}
\includegraphics[scale=0.5]{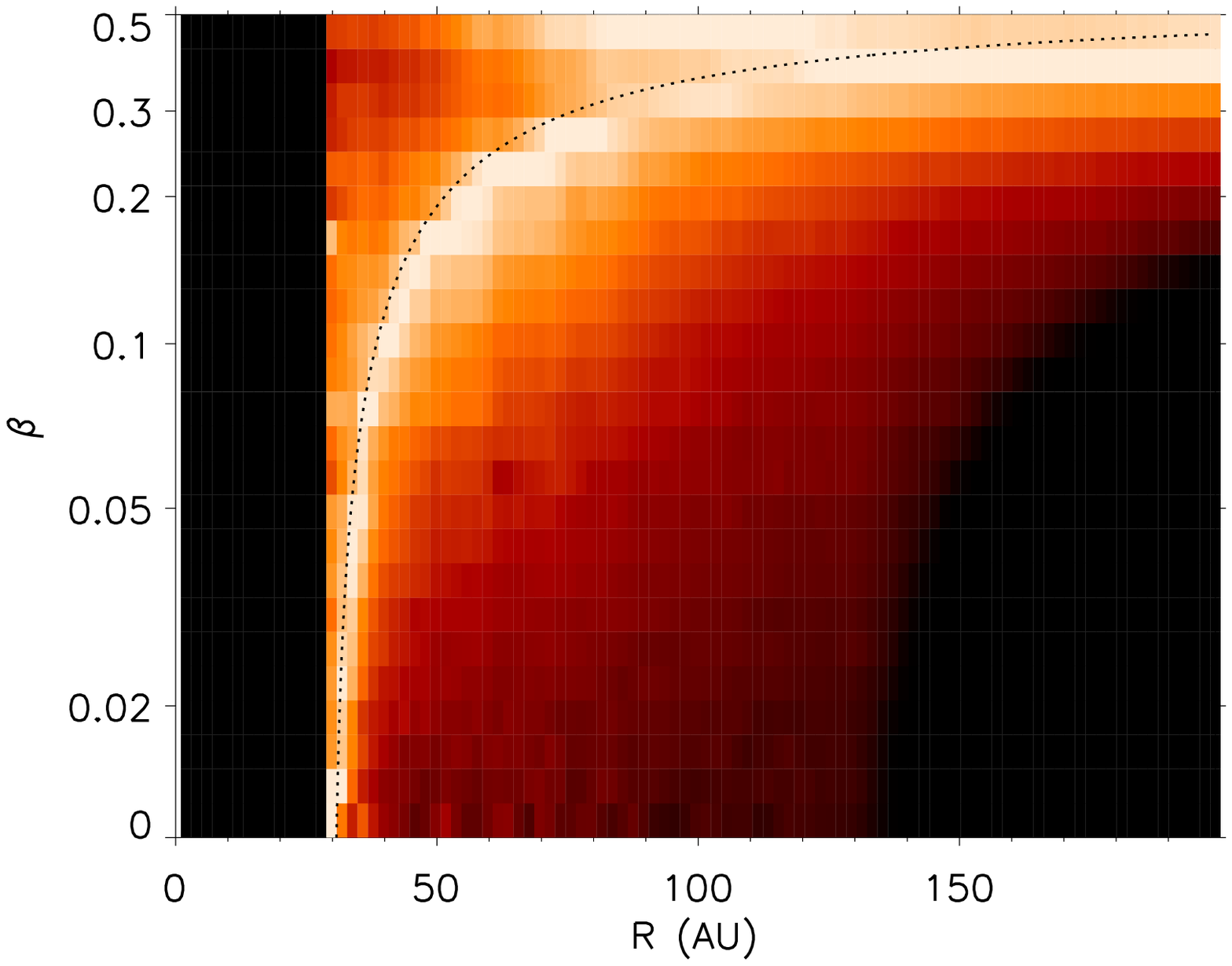}
\color{white} 
 \put(-50,40){b)} 
\color{black}
}
\caption[]{Same as Fig.\ref{intno}, but for a wide disc initially extending between 30 and 130\,AU.
}
\label{extno}
\end{figure*}

\begin{figure}
\includegraphics[width=\columnwidth]{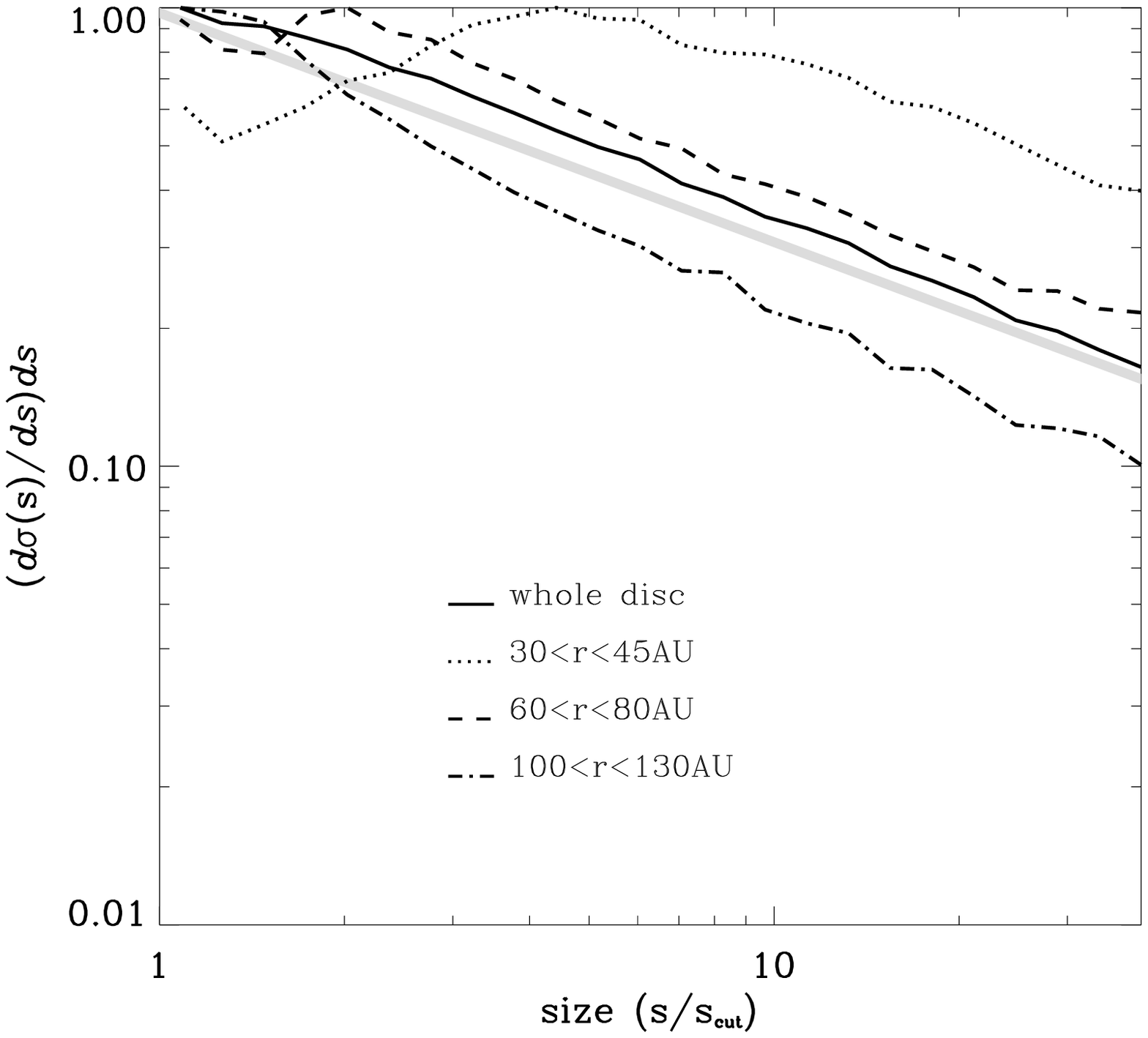}
\caption[]{Same as Fig.\ref{intnosz} but for a wide disc initially extending between 30 and 130\,AU.
}
\label{extnosz}
\end{figure}
\begin{figure}
\includegraphics[width=\columnwidth]{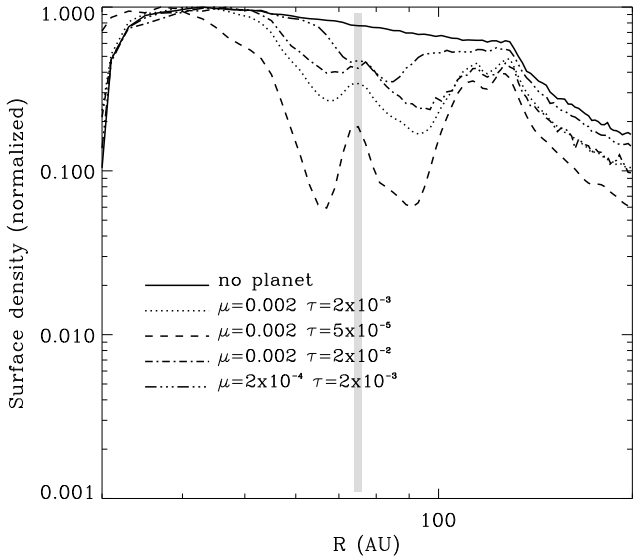}
\caption[]{Same as Fig.\ref{intnocut}, but for a wide disc initially extending between 30 and 130\,AU. The grey strip indicates the radial location of the planet for all considered runs (75AU).
}
\label{extnocut}
\end{figure}

In this case as well (run B), small dust grains populate the regions beyond the outer border of the initial parent-body disc (Fig.\ref{extno}a). The particle size distribution, although less spatially segregated than for the narrow ring case, is not homogeneous throughout the disc (Figs.\ref{extno}b and \ref{extnosz}). This is in particular true in the inner $\lesssim 70\,$AU region, which is depleted in small grains. In these regions, the geometrical optical depth is no longer dominated by the smallest bound grains, as it is in a standard PSD of index $\leq -3$. Instead, at any given radial distance $r$, the geometrical optical depth is dominated by particles produced around the inner edge of the disc and with a $\beta =\beta_Q(r)$ such that their apoastron is located at $\sim r$ (Fig.\ref{extno}b). 
Note, however, that the PSD is much less peaked around $\beta_Q(r)$ than it was beyond the narrow ring considered in the previous section. This is because in the present case, at a given location $r$ in the disc, in addition to the particles produced at the inner edge and with an apastron located at $r$, there is also a \emph{local} population of dust produced from parent bodies located there. As a consequence, the PSD stays much closer to a standard law in $s^{-3.5}$, except very close to the inner edge (compare Fig.\ref{extnosz} to Fig.\ref{intnosz}).
In fact, had we taken a flat radial density profile for the parent-body disc instead of a $\propto r^{-1}$ one, then the PSD throughout the disc would have been much closer to a standard PSD where the smallest bound grains dominate. In the present case, the dominance of grains around $\beta_Q(r)$ does indeed arise because they originate from regions that are denser then the one at radial distance $r$ (because of the $r^{-1}$ slope of the parent-body profile).

\subsection{Planet-perturbed discs}

\subsubsection{Narrow ring}

\begin{figure*}
\makebox[\textwidth]{
\includegraphics[scale=0.5]{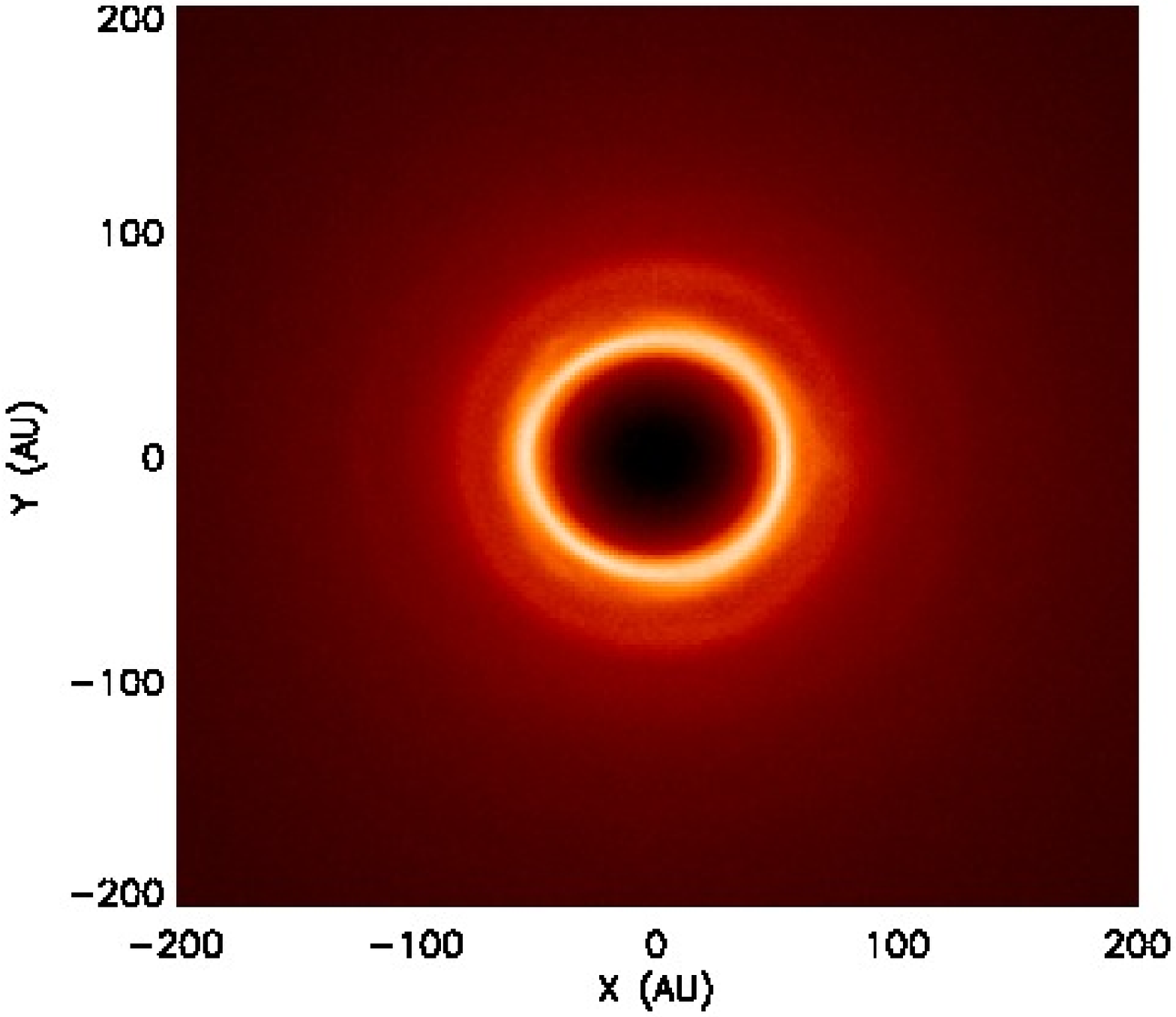}
\color{white} 
 \put(-200,40){a)} 
\color{black}
\color{white} 
 \put(-77,115){$^{\oplus}$} 
\color{black}
\includegraphics[scale=0.5]{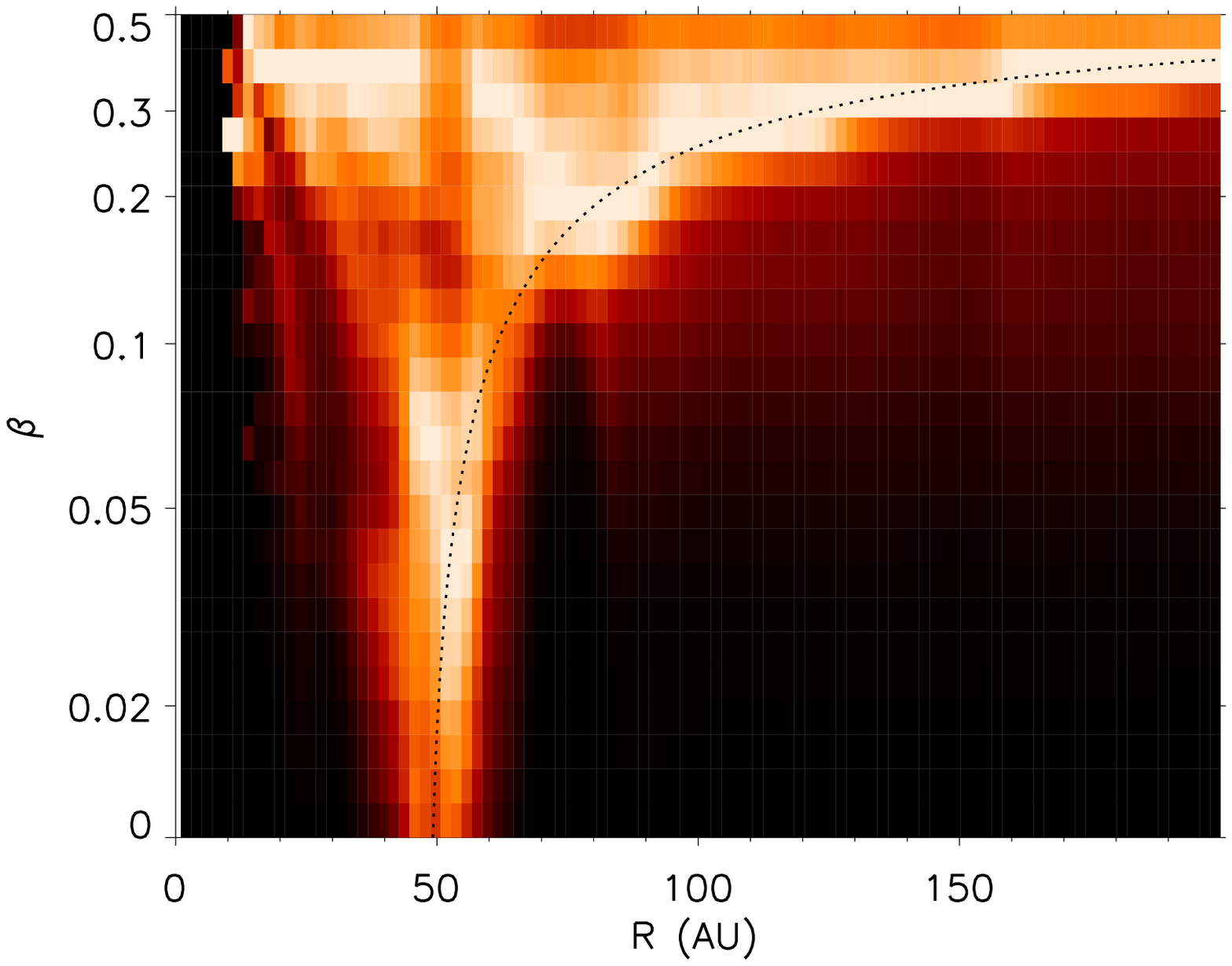}
\color{white} 
 \put(-50,40){b)} 
\color{black}
}
\makebox[\textwidth]{
\includegraphics[scale=0.5]{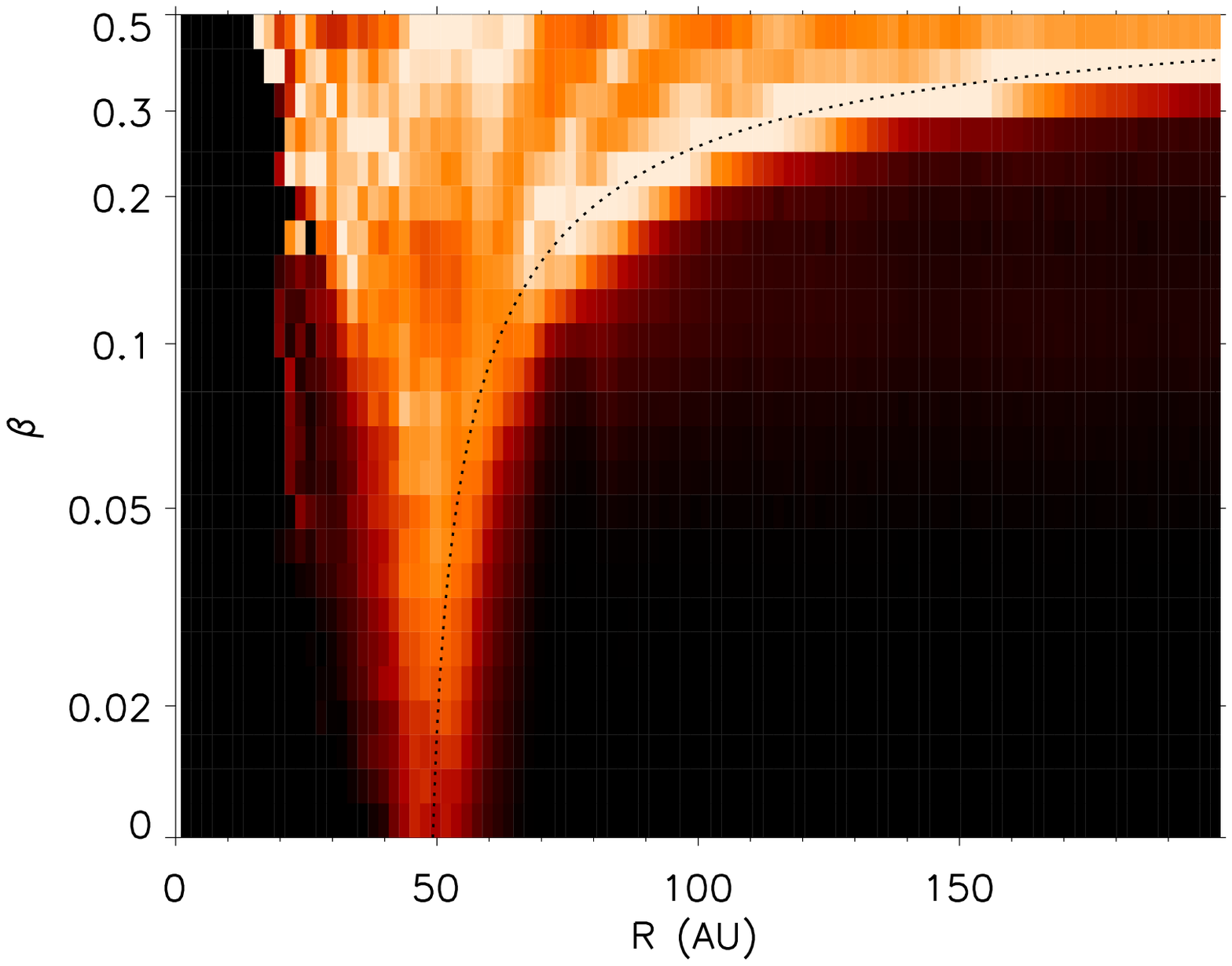}
\color{white} 
 \put(-200,40){c)} 
\color{black}
\includegraphics[scale=0.5]{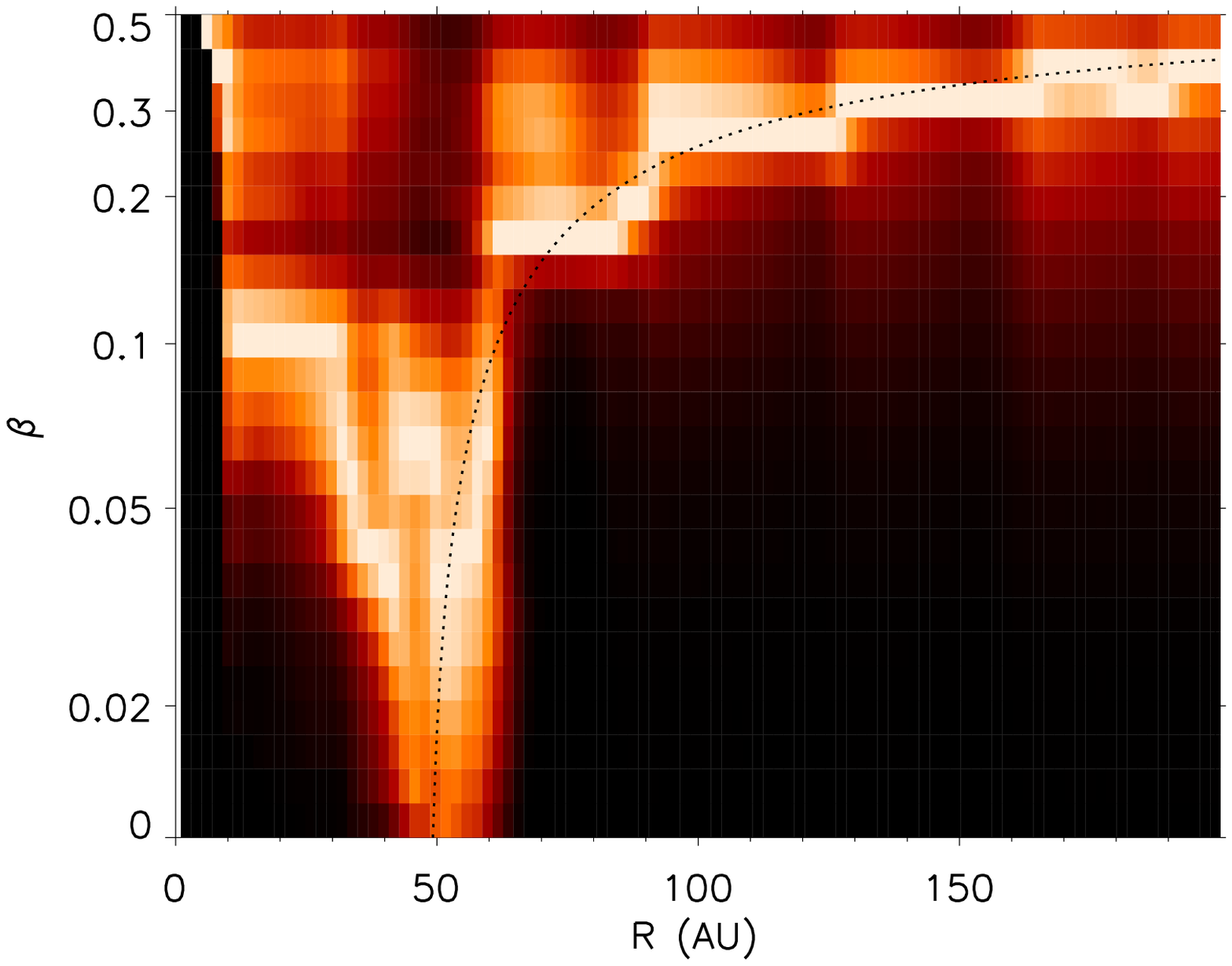}
\color{white} 
 \put(-50,40){d)} 
\color{black}
}
\caption[]{Narrow ring perturbed by an outer planet. \,\,\emph{a}): Normalized surface density profile viewed head-on, for the nominal case with $\mu=0.002$, $a_p=75\,$AU and $\tau_O=2\times 10^{-3}$ (run C). The planet's location is marked by the ${\oplus}$ symbol;  \,\,\emph{b}): Azimuthally averaged radial distribution of the geometrical cross-section as a function of particle sizes (as parameterized by their $\beta$ ratio) for run C,  \,\,\emph{c}): Same as b), but for a denser and more collisionally active disc with $\tau=2\times 10^{-2}$ (run E);  \,\,\emph{d}): Same as b), but for a more tenuous disc with $\tau=5\times 10^{-5}$ (run D).
}
\label{inte0}
\end{figure*}

\begin{figure}
\includegraphics[width=\columnwidth]{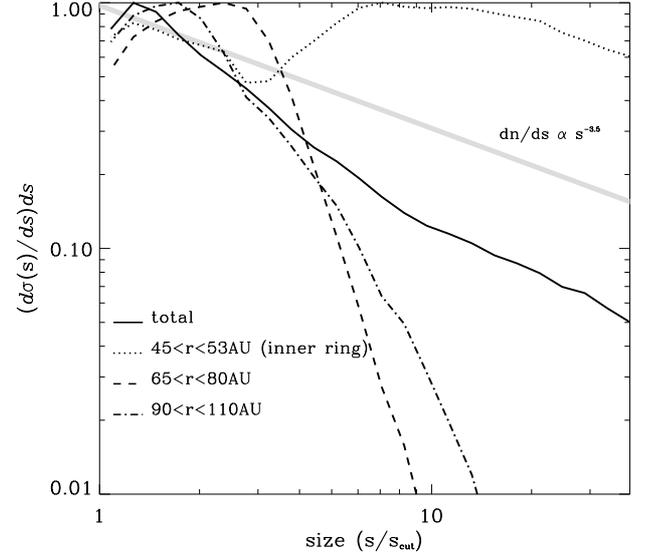}
\caption[]{Narrow ring perturbed by an outer planet. Geometrical cross section distribution per logarithmic size bin. 
}
\label{inte0sz}
\end{figure}

We now consider the same parent-body ring as in Sec.\ref{narri} but for the case where it has been sculpted by a planet exterior to it. To ensure this, the planet is placed at exactly the right distance from the ring so that it truncates it around 53\,AU. We obtained this optimal planet distance by a trial and error procedure until the correct outer truncation of the parent body ring was reached. For our nominal case of a $\mu = 0.002$ planet on a circular orbit, the corresponding planet location is 75\,AU.
We then run our series of collisional runs handling small radiation-pressure-affected grains, and obtained the steady-state displayed in Fig.\ref{inte0}. 

We retrieved the results of \citet{theb12a}, which are that the planet does not prevent the presence of matter far beyond the main ring. The density profile beyond the ring is even relatively similar to the one obtained in the case without a planet, the only difference being a slight dip at the planet's location (Fig.\ref{intnocut}).
There are, however, differences when it comes to the particle size distribution. The main parent-body ring becomes strongly depleted from small grains whose orbits intersect that of the planet at the risk of being ejected. As a result, inside the main ring it is particles with sizes $s\gtrsim 10s_{cut}$ that dominate the geometrical cross section (Figs.\ref{inte0}b and \ref{inte0sz}). Beyond the main ring, the optical depth is still roughly dominated by main-ring-produced small grains at their apastron (along the dotted curve), but the PSD is much less peaked around this size $\beta_Q(r)$ than in the case without a planet. Note, for instance, the presence of a non-negligible population of larger grains ($0.05 \lesssim \beta \lesssim 0.2$) beyond the planet's location, in regions that radiation pressure alone does not allow them to reach. These grains populate these regions because of the planet's perturbations, albeit on a temporary basis, but long enough to contribute to the optical depth there.

As could be logically expected, when increasing the collisional activity within the ring (by increasing the scaling factor $\tau_0$ to $2\times 10^{-2}$, run E) the shielding and sculpting effect of the planet becomes less pronounced. While the global surface density profile is relatively similar to the reference $\tau_0=2\times 10^{-3}$ case (Fig.\ref{intnocut}), the PSD becomes much closer to that of a planet-free case, with, for instance, small grains again dominating in the parent-body ring (Fig.\ref{inte0}c). Conversely, for more tenuous and less collisional discs ($\tau_0=5\times10^{-5}$, run D), the planet's truncating effect is stronger: there is much less matter beyond the main ring and the radial density profile is much steeper (Fig.\ref{intnocut}), this time strongly departing from the profile without a planet. Likewise, the depletion of small grains within the ring is much more pronounced (Fig.\ref{inte0}d), because they are much faster removed by interactions with the planet than they are collisionally produced.

\subsubsection{Extended disc}

\begin{figure*}
\makebox[\textwidth]{
\includegraphics[scale=0.5]{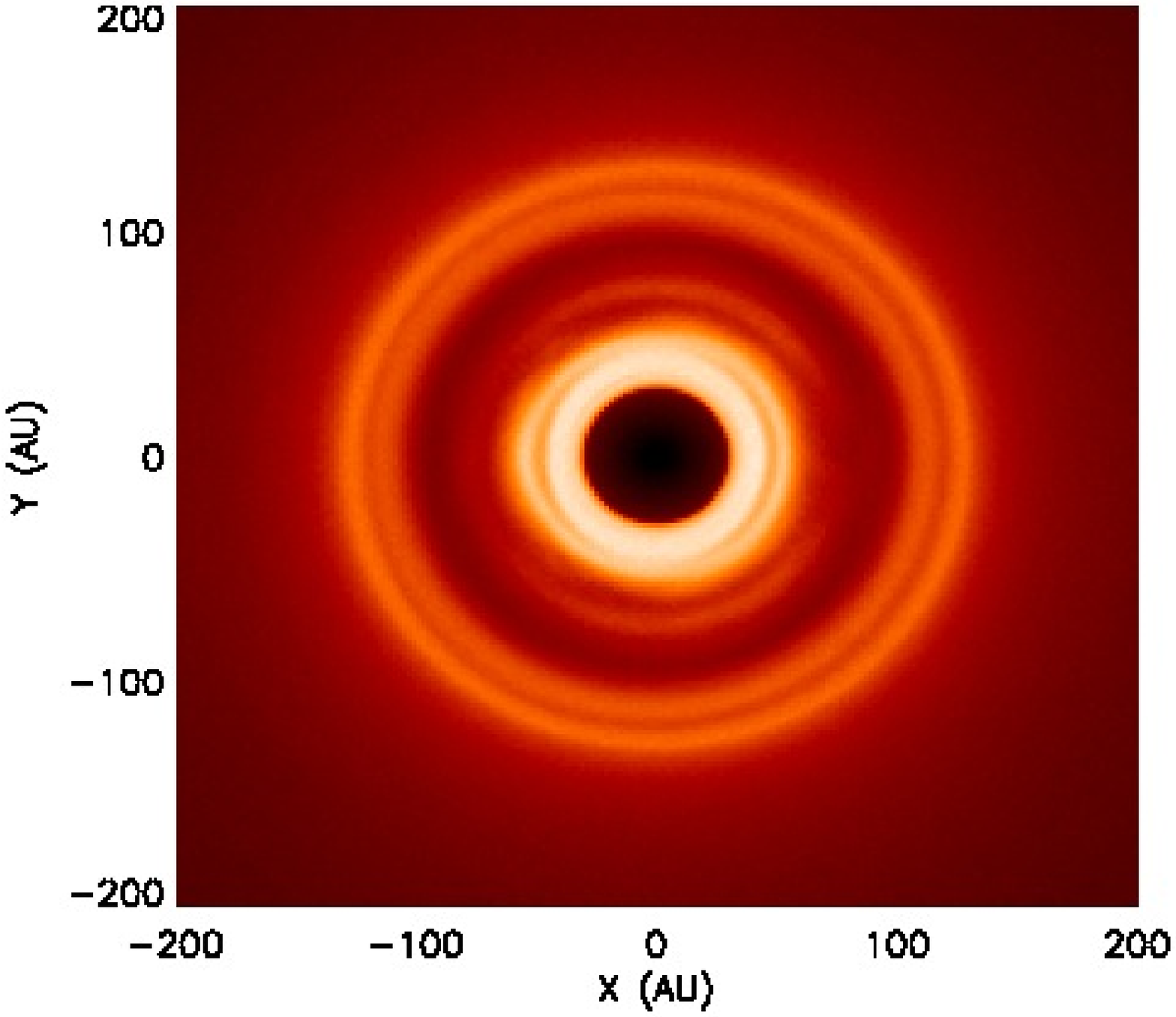}
\color{white} 
 \put(-200,40){a)} 
\color{white}
 \put(-77,115){$^{\oplus}$} 
\color{black}
\includegraphics[scale=0.5]{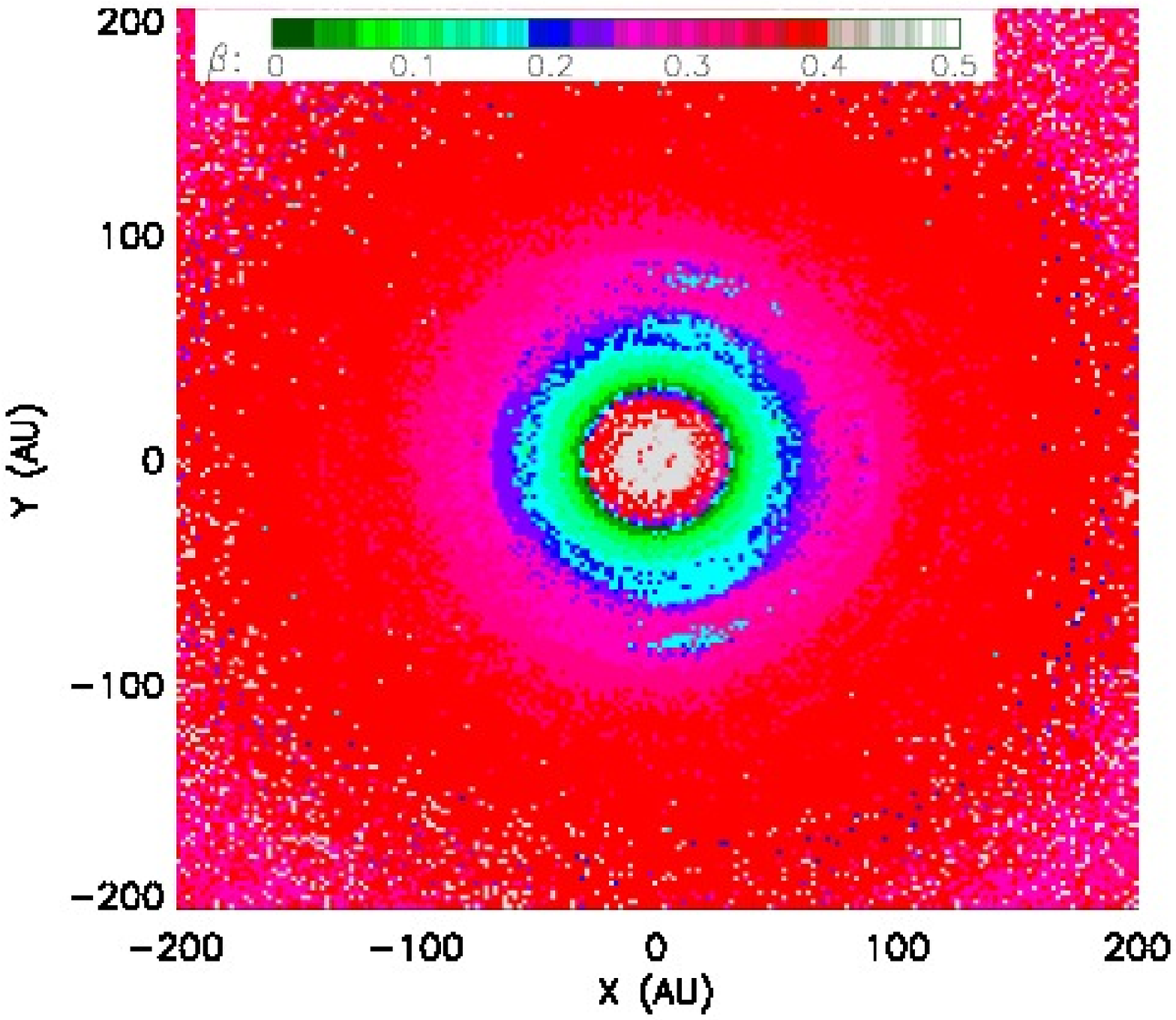}
\color{black} 
 \put(-200,40){b)} 
\color{black}
 \put(-77,115){$^{\oplus}$} 
\color{black}
}
\caption[]{Extended disc perturbed by an embedded planet. \emph{a}): Normalized surface density profile viewed head-on; \emph{b}): particle size (as parameterized by the $\beta$ ratio) dominating the geometrical cross section. The planet has a mass $\mu=0.002$ and is on a circular orbit at $75\,$AU (nominal run G with $\tau_0 = 2\times 10^{-3}$).
}
\label{widee0}
\end{figure*}

\begin{figure*}
\makebox[\textwidth]{
\includegraphics[scale=0.5]{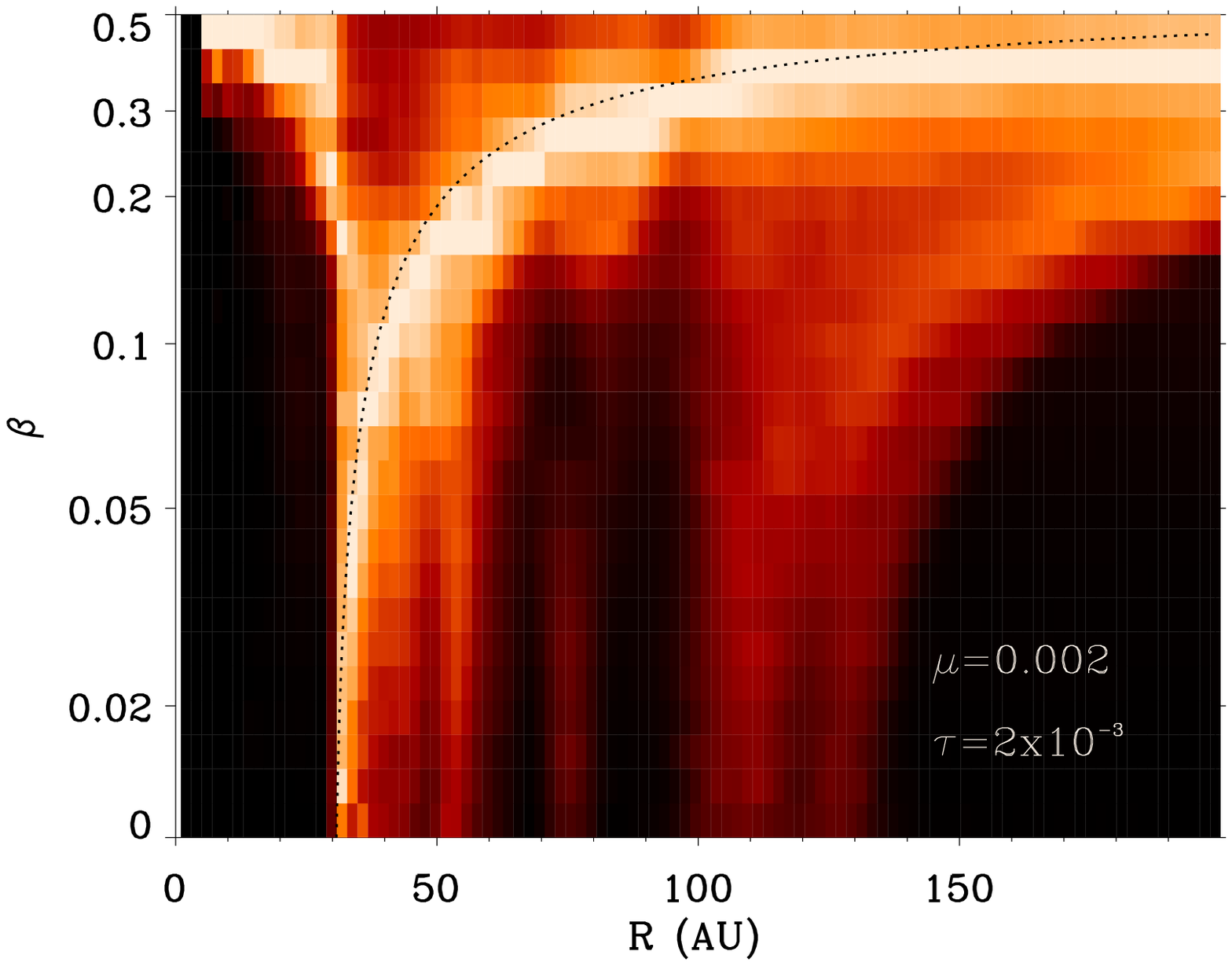}
\color{white} 
 \put(-200,40){a)} 
\color{black}
\includegraphics[scale=0.5]{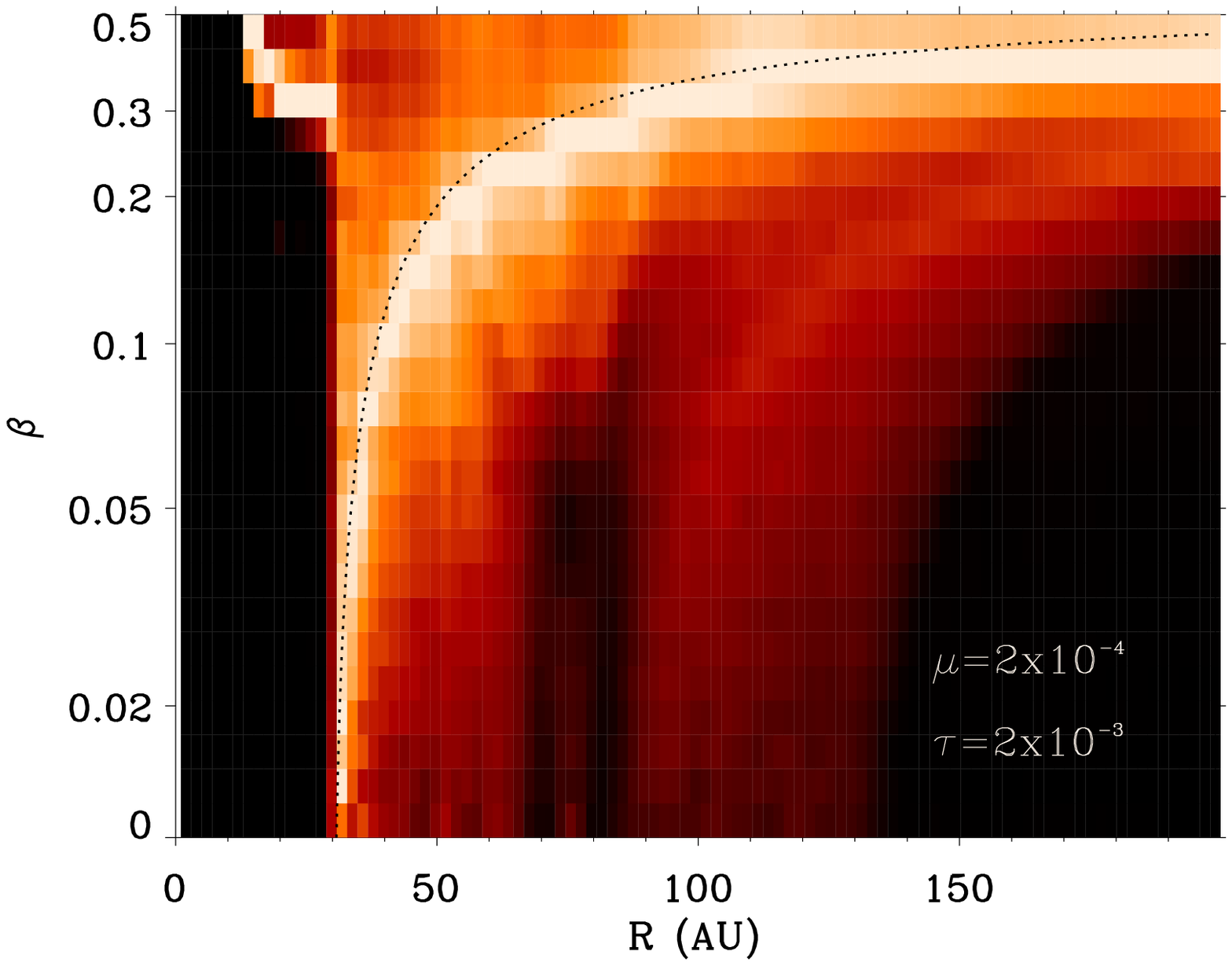}
\color{white} 
 \put(-200,40){b)} 
\color{black}
}
\makebox[\textwidth]{
\includegraphics[scale=0.5]{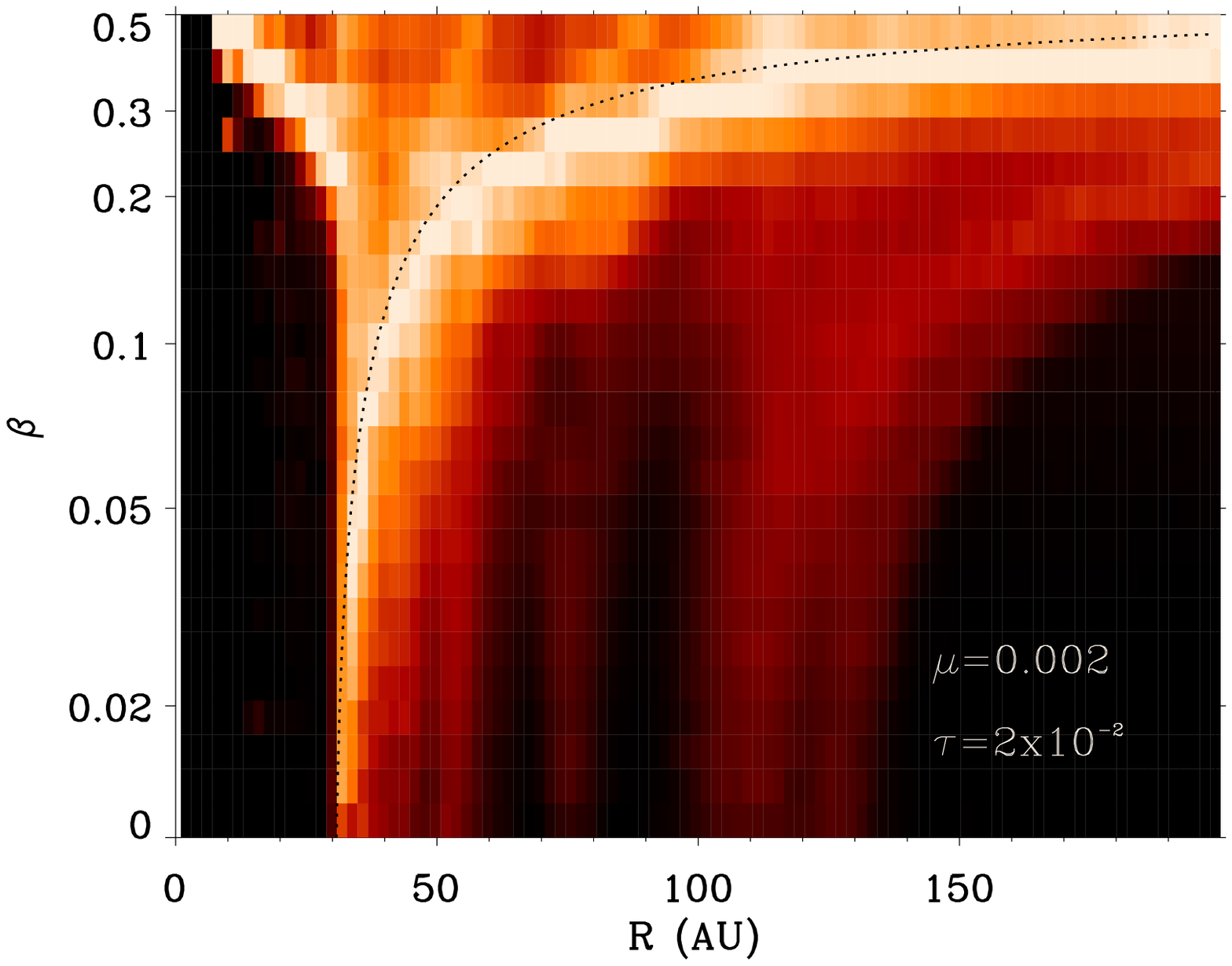}
\color{white} 
 \put(-200,40){c)} 
\color{black}
\includegraphics[scale=0.5]{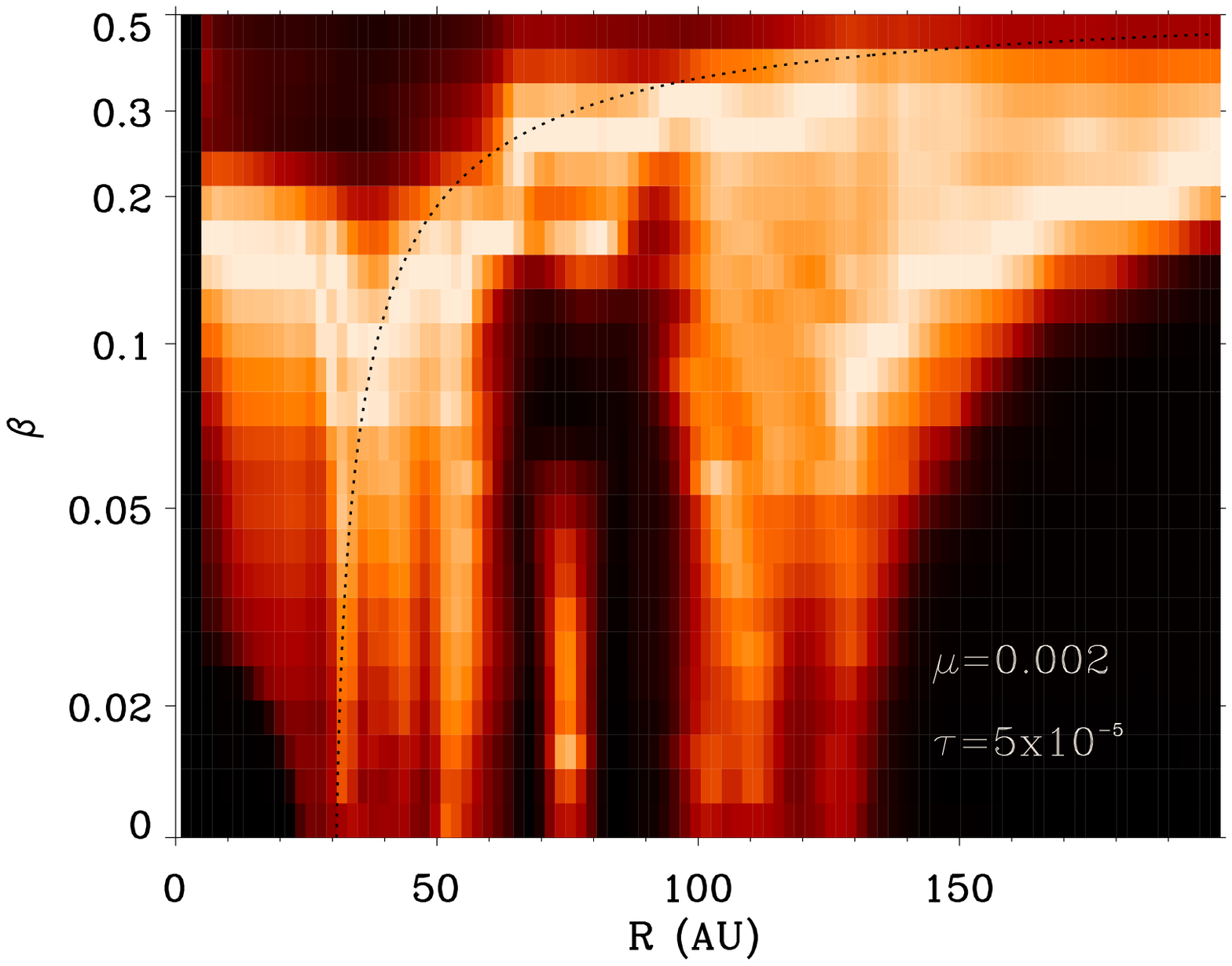}
\color{white} 
 \put(-200,40){d)} 
\color{black}
}
\caption[]{Extended disc perturbed by an embedded planet. Azimuthally averaged radial distribution of the geometrical cross-section, as a function of particle sizes, for four different configurations: a) Nominal (run G), b) Saturn-mass planet ($\mu = 2\times 10^{-4}$, run J), c) high optical depth ($\tau=2\times10^{-2}$, run I), d) low optical depth ($\tau=5\times10^{-5}$, run H).
}
\label{e0comp}
\end{figure*}

\begin{figure}
\includegraphics[width=\columnwidth]{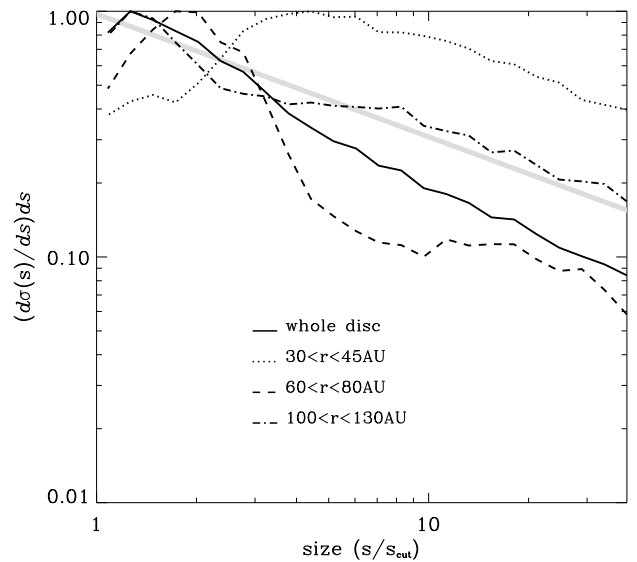}
\caption[]{Extended disc perturbed by an embedded planet. Geometrical cross section distribution per logarithmic size bin. 
}
\label{e0sz}
\end{figure}

For the extended-disc case, the perturbing planet is embedded within the disc, on a circular orbit at 75\,AU. The expected purely dynamical structures are clearly visible (Fig.\ref{widee0}a): a gap surrounding the planet's radial location, local overdensities at the two stable corotating Lagrangian points L4 and L5, and density gaps at the interior and exterior 1:2 and 2:1 resonances, at 47 and 119\,AU, respectively. However, as shown by \citet{theb12b}, the combined effect of collisions and radiation pressure makes this picture much more complex. As an example, the dynamically unstable gap region is not empty but populated by small grains produced farther in. Likewise, almost the entire corotating region is populated by small grains, instead of them being clustered around the Lagrangian points. 

The situation is even more complex for the particle size distribution (Figs.\ref{widee0}b and \ref{e0comp}). The shielding effect of the planet, cutting the eccentric orbits of small grains, amplifies the depletion of small grains in the inner regions. However, this shielding effect is not enough to prevent these inner-region-produced small grains from dominating the geometrical cross section in the outer regions of the disc (beyond the planet's orbit). The gap region around the planet's location is almost devoid of large grains (with $\beta \lesssim 0.1$), but not of smaller particles (Fig.\ref{e0sz}). An important exception is the close vicinity of the two Lagrangian points, where mostly large grains are present (Fig.\ref{widee0}).

We explore in Fig.\ref{e0comp} how the PSD varies for different initial setups. The imprint of a sub-Saturn-mass planet (run J) on the PSD is still visible in the PSD, but is restricted to the close vicinity of the planet while being close to the case without a planet in the rest of the disc. Varying the collisional activity of the disc (by tuning in $\tau_0$) has significant consequences as well. For low values of $\tau_0$ (run H), the effect of the planet on the PSD is much more pronounced, notably because the rate at which small high-$e$ particles are produced in the inner disc is not enough to balance the rate at which they are dynamically removed by close encounters with the planet. As a consequence, the $r\leq 60\,$AU region is almost devoid of $\beta \geq 0.2$ grains. This has a direct effect on the disc's surface density, for which the density drop in the gap region around the planet is much more pronounced than in the case with $\tau_0=2\times 10^{-3}$, as is the peak due to large grains corotating with the planet (see Fig.\ref{extnocut}). Another consequence is that fewer inner-region-produced grains are present in the disc's outer regions (beyond $\sim 90\,$AU), where the contribution of locally produced larger grains becomes predominant. Moreover, the depletion of large grains in the 1:2 or 2:1 resonances is now clearly visible in the radial distribution of the PSD whereas it was barely noticeable in the nominal $\tau_0=2\times 10^{-3}$ case. 
As expected, for a highly collisional disc ($\tau_0=2\times 10^{-2}$, run I), the PSD approaches that of the case without a planet, that is, inner-regions-produced grains dominating the whole disc and a much larger number of small grains present in the $\leq 60\,$AU region. Conversely, the density gap around the planet is shallower than in the nominal case, even though the planet still leaves a clear signature when the system is seen head-on (see \citet{theb12b} for a discussion on planet signatures for systems seen edge-on).

\subsection{Synthetic images}

\begin{figure*}
\makebox[\textwidth]{
\includegraphics[scale=0.5]{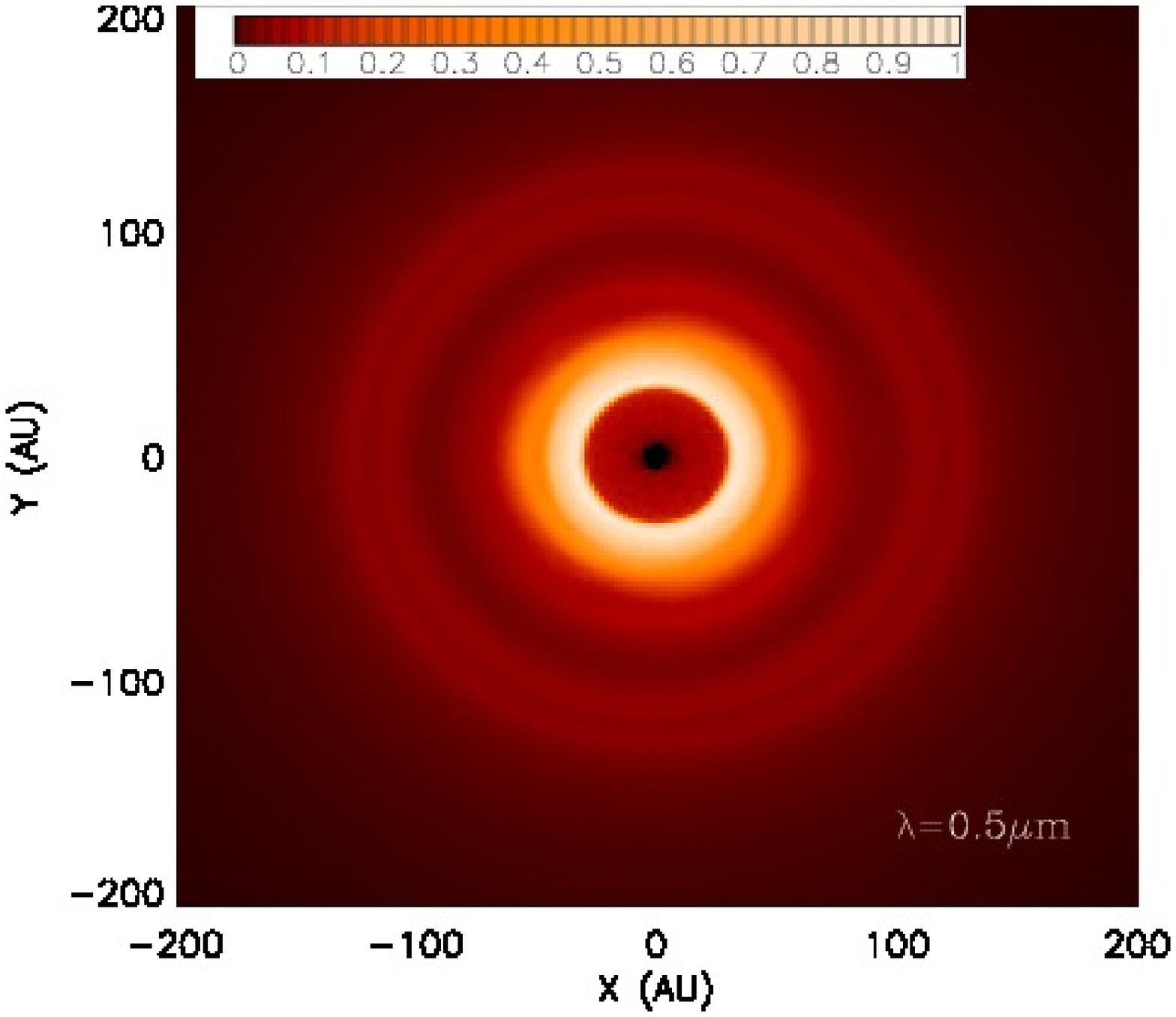}
\color{white} 
 \put(-200,40){a)} 
\color{black}
\includegraphics[scale=0.5]{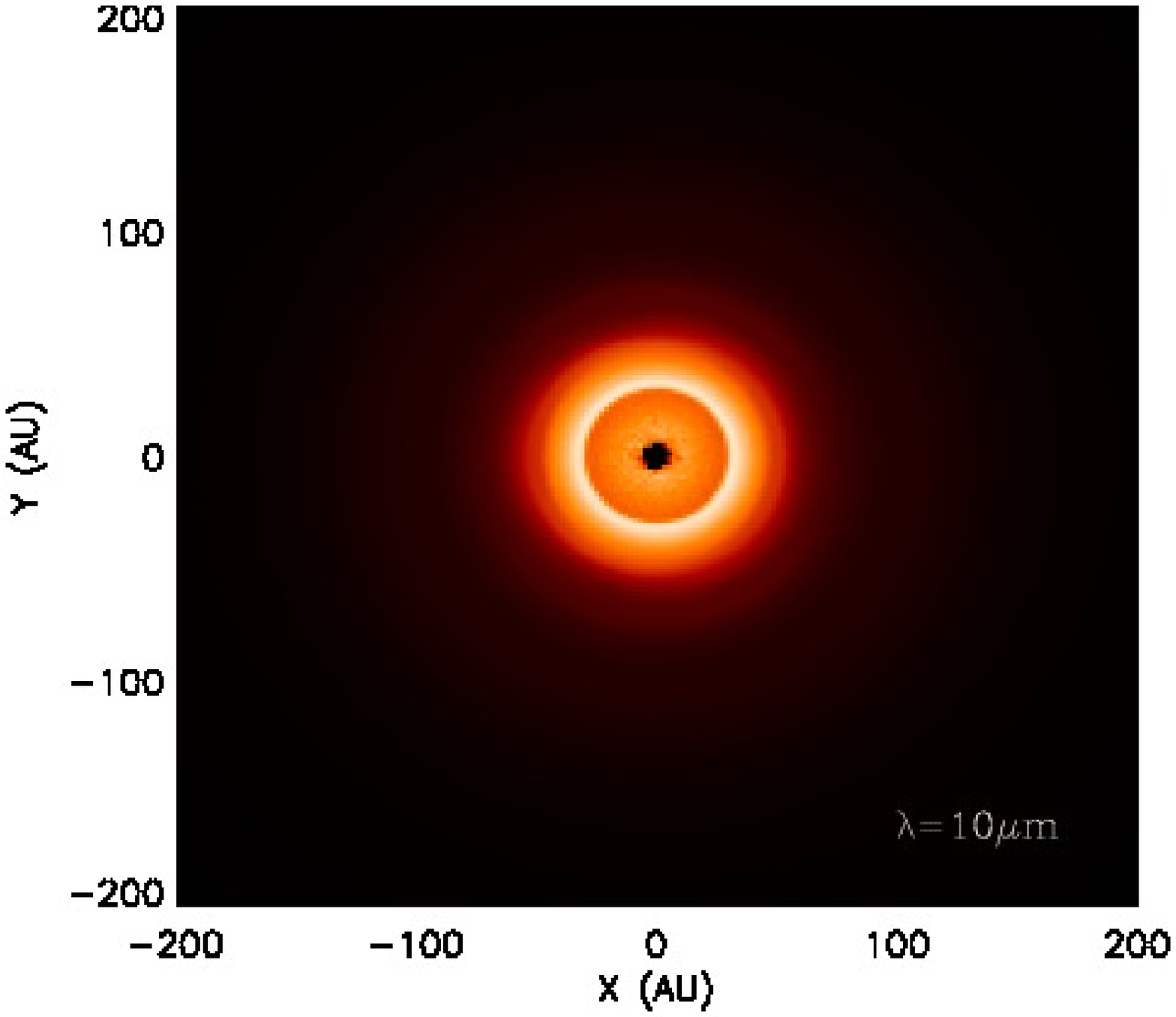}
\color{white} 
 \put(-200,40){b)} 
\color{black}
}
\makebox[\textwidth]{
\includegraphics[scale=0.5]{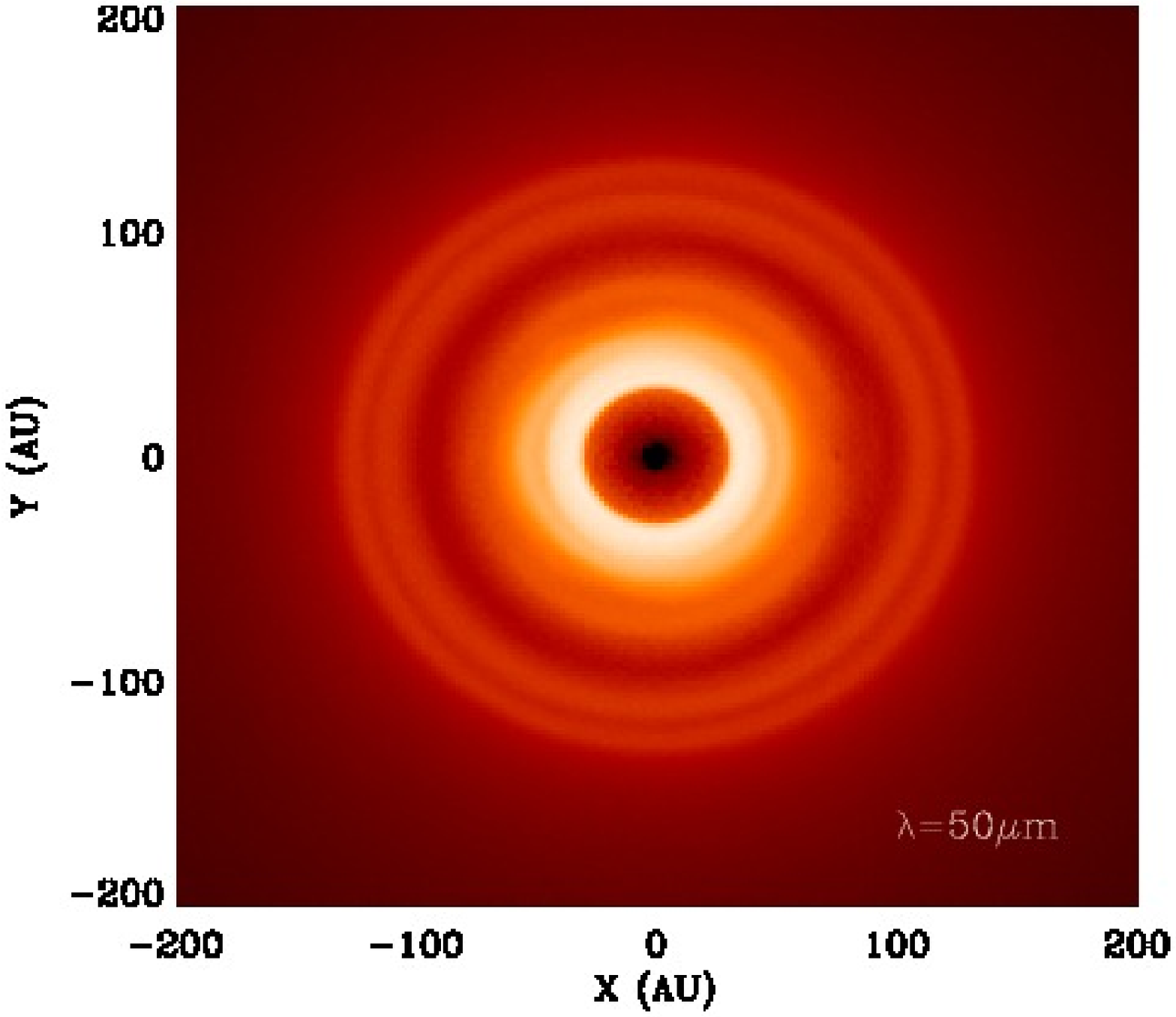}
\color{white} 
 \put(-200,40){c)} 
\color{black}
\includegraphics[scale=0.5]{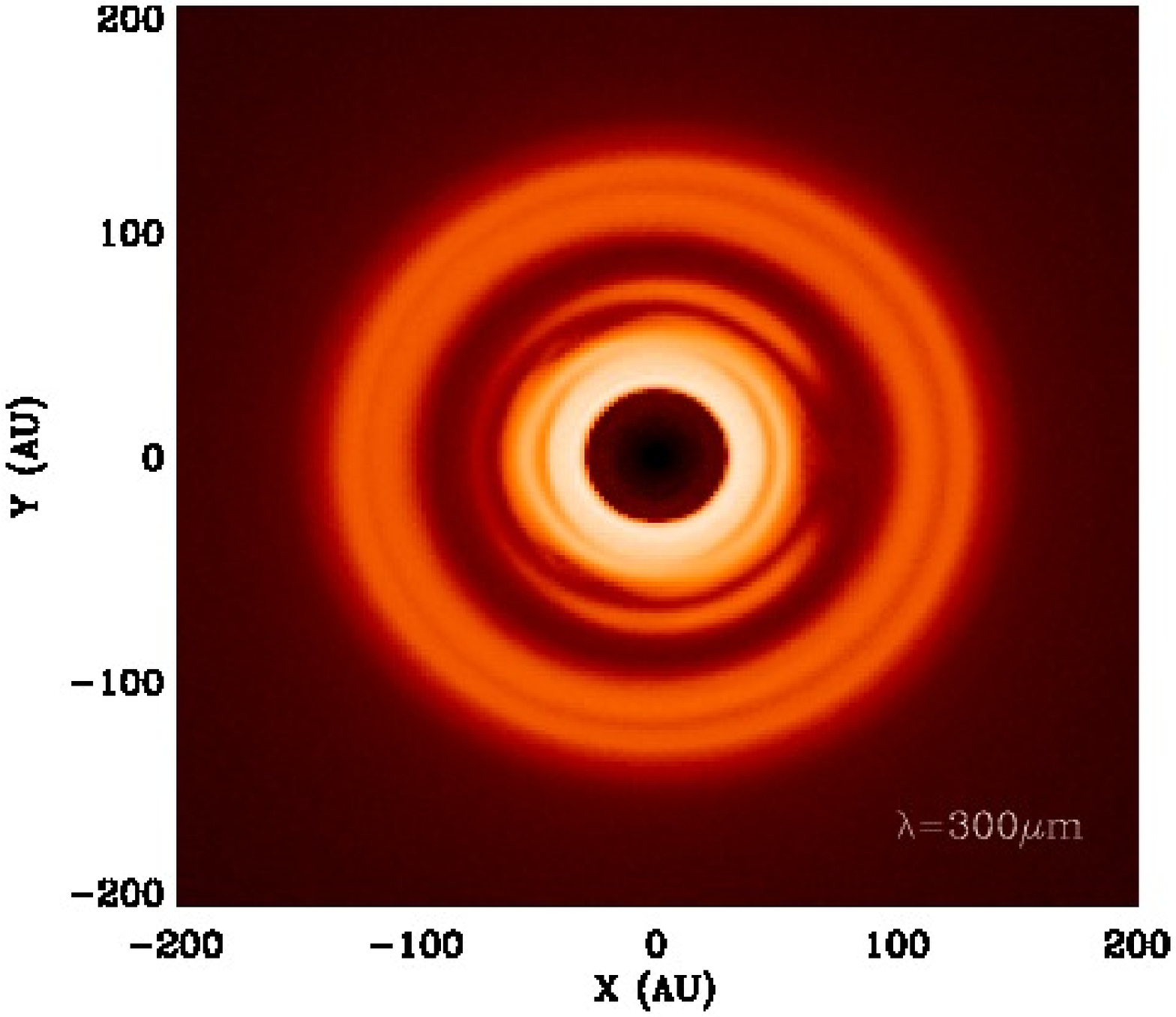}
\color{white} 
 \put(-200,40){d)} 
\color{black}
}
\caption[]{Synthetic images produced with GRaTeR for the extended-disc/embedded-planet case (nominal case, run G) assuming the central star is an A0V. We consider four different wavelengths: \emph{a}): $0.5\mu$m (scattered light); \emph{b}): $10\mu$m; \emph{c}): $50\mu$m and \emph{d}): $300\mu$m. In each image, the flux has been normalized.
}
\label{immu}
\end{figure*}

To investigate to what extent spatially segregated PSDs can affect images of resolved discs, we present in Fig.\ref{immu} synthetic images derived for our nominal case of a wide disc with a planet (run G). To derive these images, an additional parameter needs to be fixed, that is, the spectral type of the star (whereas this parameter was unconstrained for our surface density maps and PSD profiles, for which all grain sizes had been rescaled by their $\beta$ ratio). We chose here a bright A star, typical for stars around which the best resolved debris disc images have been obtained (i.e., $\beta$ Pic, Vega, etc.). For the chosen A0V stellar type, the radiation-pressure blow-out size $s_{cut}$ is $\sim 5\mu$m for compact silicates. After fixing the stellar type, images are produced by plugging our particle size and spatial distributions into the GRaTeR package \citep{auge99}. 

Fig.\ref{immu} shows that the aspect of our disc+planet system drastically changes depending on the wavelength at which it is observed. This is of course an expected result, especially in thermal emission, where dust temperature (and thus its size and radial location) plays a crucial role. This is, for example, the main cause for the important difference between the image in scattered light (Fig.\ref{immu}a), which does not depend on grain temperature, and the one at $10\mu$m (Fig.\ref{immu}b), which does and is only weakly sensitive to the cold grains beyond the planet's location at 75\,AU. This temperature effect is also responsible for the fact that outer regions become brighter for increasing wavelengths.

However, in addition to this temperature effect, the disc's aspect as a function of wavelength is also impacted by the spatially varying size distribution. One important point is that observations at a given wavelength $\lambda$ are typically weakly sensitive to grains $\leq\lambda/2\pi$. This will have observable consequences for $\lambda$ longwards of $2\pi s_{cut} \sim 30\mu$m. A clear illustration of this is that the image at $300\mu$m has much sharper structures than the one at $50\mu$m, mainly because a $300\mu$m image does not display small grains with $\beta \gtrsim 0.5\times s_{cut}/(300/2\pi) \sim 0.1$. Since these grains are the ones that populate the regions beyond the parent-body disc's outer edge as well as the unstable regions in the planet-induced gap, it follows that the edges of the gap, the coorbital stable Lagrangian populations and the disc's outer edge are much more clearly defined than at shorter wavelengths\footnote{Note that we are here only considering astrophysical effects and do not take into account the resolution of the observing instruments, which is usually better at shorter wavelengths.}.
Another consequence of the segregated PSD in the $300\mu$m image is that the contrast between the inner and outer regions of the disc is still largely in favour of the former, even though they are too hot to peak at such long wavelengths. This is because the outer regions are dominated by tiny grains with $\beta\gtrsim0.3$ (see Fig.\ref{e0comp}), and thus sizes $\leq\lambda/2\pi$, which are poor emitters at $300\mu$m, while the inner regions are dominated by larger grains ($\beta\lesssim0.1$), which emit more efficiently at such long wavelengths.

\section{Discussion and conclusions}

Our numerical investigation has shown that, as a result of the coupling between collisions, radiation pressure, and possible dynamical perturbations, the size distribution of debris disc particles is always significantly spatially segregated, at least for stars massive enough (typically $\geq 0.9M_{\odot}$), for $\beta$ to reach values higher than 0.5. For all explored set-ups, the only case for which we obtain a PSD that is close to a standard power-law in $s^{-3.5}$ is that of an unperturbed narrow ring of parent bodies. But even for this case, the standard power-law was only obtained inside the narrow ring itself, the matter beyond the ring being strongly size-segregated. For all the other cases, that is, extended disc at rest or with an embedded planet, narrow ring with perturbing planet, etc., we were unable to find a region of the discs where the PSD follows such a standard power-law. 
The amplitude of these departures from standard PSDs vary, in a relatively complex way, with the considered set-up and with spatial location within the disc. Because of these complex behaviours, it is difficult to derive generic rules for the behaviour of PSDs in a given environment. However, some general reliable results can be given:
\begin{itemize}
\item Contrary to the prediction of standard PSDs in $s^{q}$ (with $q\leq -3$), we find that the size dominating the geometrical cross section is in many cases \emph{not} that of the smallest bound particles. Cases for which larger grains dominate are notably, \emph{a)} the inner regions of extended discs (see Figs.\ref{extnocut} and \ref{e0comp}), \emph{b)} the regions just beyond a narrow ring (Fig.\ref{intno}), \emph{c)} the regions radially below a perturbing planet's orbit (Figs.\ref{inte0} and \ref{e0comp}), and \emph{d)} the corotating L4 and L5 Lagragian points around disc-embedded planets (Fig.\ref{widee0}). The dominance of large grains in these regions is more pronounced for tenuous discs than for dense ones.
\item For an unperturbed narrow ring of parent bodies, a standard PSD holds \emph{within} the narrow ring itself. Beyond the ring, the size distribution is peaked around a narrow size range. At a radial distance $r$ beyond the ring's location $r_0$, the size $s^{*}$ that dominates is given by $r_0=(1-2\beta(s^{*}))\times r$, corresponding to grains produced in the ring and placed by radiation pressure on an orbit with a apastron located at $r$.
\item For an unperturbed extended disc, under the reasonable assumption that the surface density $\Sigma(r)$ of large parent bodies decreases with $r$, we found that, at a radial distance $r$ the PSD assumes a $dn \propto s^{q}ds$ profile down to a size $s^{*}$ given by $r_{in}=(1-2\beta(s^{*}))\times r$, where $r_{in}$ is the inner edge of the parent-body disc. The magnitude of the depletion of $s\leq s^{*}$ grains depends on the slope of the $\Sigma(r)$ profile. If the disc is wide enough, then the PSD approaches a standard power-law in the outer regions of the disc.
\item For all planet-perturbed discs, dynamically unstable regions are not empty, because of the combined effects of collisions and radiation pressure. Their PSDs are, however, very peculiar, because they are mostly populated by very small grains with typically $\beta \geq 0.15$.
\item Segregated PSDs affect the way resolved discs appear at different wavelengths. The relative luminosities between the different regions of a disc can differ from what they should be if the PSD were homogeneous. As an example, at long wavelengths $\lambda_{FIR}$, the inner regions of an extended disc can appear to be much brighter than the outer ones, even though grains in these inner regions are too hot to peak at $\lambda_{FIR}$. This is because the outer regions are populated by smaller grains, which ar poor emitters at long wavelengths despite having a more "adequate" temperature.
Likewise, planet-induced structures are expected to appear sharper at longer wavelengths and more blurry at shorter ones because unstable regions are predominantly populated by smaller grains than stable regions are.
\end{itemize}

Some of these robust results are simple enough to be useful for image- and SED-fitting procedures. As an example, our results for the specific case of unperturbed systems (narrow ring or extended discs) can be easily implemented into best-fit models such as GRaTeR. A possible method would be to divide the system to be fitted into concentric radial annuli, which would be fitted sequentially, following an inside-out direction. The initial step would be to find the best PSD in the innermost annulus (at distance $r_0$) using a standard fitting procedure. From this PSD, one is then able to deduce the number of small grains, having high-$e$ orbits due to radiation pressure, which are needed in the outer annuli (with $r\geq r_0$) in order to obtain the best-fit PSD of the $r_0$ annulus \footnote{Note that this procedure is different from that, used in many past studies, which simply dilutes the number of small grains along their elongated orbit, thus resulting in an underabundance of these grains in the inner annulus.}. This can be done by solving Kepler's equation when assuming that these grains have their periastron at $r_0$ and an orbital eccentricity $e=(1-\beta)/(1-2\beta)$. For the next radial annulus at $r_1$, the same procedure needs to be applied, except that one needs to take into account in the fit the number of high-$e$ grains coming from the $r_0$ annulus. Once a best PSD is obtained at $r_1$, one switches to the next radial annulus and the procedure is iterated until the outer edge of the system is reached.

For systems where a planetary perturber is expected (for instance because of asymmetries in resolved images), such simple methods cannot be applied. However, if a possible orbit and mass for the planet are considered, our results suggest that the PSD fitting should be divided into three separate regions: 1) the dynamically stable region below the planet's orbit, 2) the stable region beyond its orbit, and 3) the unstable region around the planet. For tenuous discs, we expect region 1 to be devoid of grains with $\beta \gtrsim 0.15$ and region 2 to have an almost unperturbed PSD, while region 3 is probably almost empty except for the Lagrangian points, which are expected to be populated by $\beta \lesssim 0.15$ grains. The situation is more complex for bright dense discs, which showed a complex interplay between grains from the inner, outer, and unstable regions (see Fig.\ref{e0comp}a and d). The only reliable characteristics is here that the unperturbed region is $not$ empty, but populated mostly by $\beta\geq 0.15$ grains.
 
We expect the implementation of these reliable results and prescriptions into best-fit models to greatly help our understanding of several specific debris discs. The best candidates for these investigations are discs that display well-defined spatial structures for which multi-wavelength observations are available. Obvious candidates are the narrow ring around HR4796A, which might (or might not) be confined by planetary perturbers \citep[e.g.,][]{lagr12}, or the Fomalhaut system, where the relation between the observed narrow annulus and the recently detected "planet-related" structure is not fully understood \citep{bole12, kala13}. Observations of these systems at different wavelengths could be tested against numerical predictions on their Particle Size Distributions to confirm or rule out possible scenarios: shepherding planets, confining inner planet, unperturbed disc, etc. In this respect the results from the ALMA interferometer (especially once it will have longer baselines) will be of crucial importance, because they give images of unprecedented resolution in the millimetre/submillimetre that allow unique access to the spatial distribution of large grains. For the spatial distribution of smaller grains, a quantum leap in resolution should be achieved with the high-resolution images, in the visible and near-IR, expected from the JWST or the SPHERE instrument on the VLT.
The detailed study of specific astrophysical cases exceeds the scope of the present work and will be the purpose of a forthcoming paper.

\begin{acknowledgements}

Q.K. is funded by the French National Research Agency (ANR) through contract ANR-2010 BLAN-0505-01 (EXOZODI). P.T. and J-C. A. acknowledge financial support by the same contract.

\end{acknowledgements}

{}
\clearpage

\end{document}